\documentclass[5p]{elsarticle}
\usepackage{url} 
\usepackage{setspace} 
\usepackage{lineno,hyperref} 
\usepackage{enumitem}
\journal{Astronomy and Computing}

\usepackage[numbers]{natbib}
\usepackage{float}

\usepackage{amsmath,amsfonts,amssymb}
\usepackage{graphicx}
\usepackage{setspace}
\usepackage{tocloft}
\usepackage{lineno}
\usepackage{multirow}
\usepackage{epsfig}
\usepackage{caption}
\usepackage{subcaption}

\usepackage{comment}

\cftpagenumbersoff{figure}
\cftpagenumbersoff{table}

\begin{document}

\title{DarpanX: A Python Package for Modeling X-ray Reflectivity of Multilayer Mirrors}                 

\author[a,b]{Biswajit Mondal}
\ead{biswajitm@prl.res.in}
\author[a]{Santosh V. Vadawale}
\author[a]{N.P.S. Mithun}
\author[a]{C.S. Vaishnava}
\author[a]{Neeraj K. Tiwari}
\author[a]{S.K. Goyal}
\author[c]{Singam S. Panini}
\author[d]{Vinita Navalkar}
\author[e]{Chiranjit Karmakar}
\author[e]{Mansukhlal R. Patel}
\author[e]{R.B. Upadhyay}

\address[a]{Physical Research Laboratory, Ahmedabad 380009, India}
\address[b]{IIT Gandhinagar, Gandhinagar 382355, India}
\address[c]{Indian Institute of Astrophysics, Bengaluru 560034, India}
\address[d]{Centre for Excellence in Basic Sciences, Mumbai 400098, India}
\address[e]{Space Applications Centre, Ahmedabad 380015, India }

\providecommand{\keywords}[1]
{
  {\textit{Keywords :}} #1
}

\begin{abstract}
Multilayer X-ray mirrors consist of a coating of a large number of alternate 
layers of high Z and low Z materials with a typical thickness of 10 -- 100 
${\textup{\AA}}$,
on a suitable substrate. Such coatings play an important role in enhancing the reflectivity of X-ray mirrors by allowing reflections at angles much larger than the critical angle of X-ray reflection for the given materials. 
Coating with an equal thickness of each bilayer (constant period multilayers) enhances the reflectivity at discrete energies, satisfying Bragg condition for the given thickness. 
However, by systematically varying the bilayer thickness in the multilayer stack (depth graded multilayers), it is possible to design X-ray mirrors having enhanced reflectivity over a broad energy range. One of the most important applications of such a depth graded multilayer mirror is to realize hard X-ray telescopes for astronomical purposes. Design of such multilayer X-ray mirrors and their 
characterization with X-ray reflectivity measurements require appropriate 
software tools that can compute X-ray reflectivity for the given set of 
parameters and geometry. We have initiated the development of hard X-ray optics for future Indian X-ray astronomical missions, and in this context, we have developed a program, DarpanX, to calculate X-ray reflectivity for single and multilayer mirrors. 
It can be used as a stand-alone tool for designing multilayer mirrors with required characteristics. But more importantly, 
it has been implemented as a local model for the popular X-ray spectral fitting program, XSPEC, and thus can be used for accurate fitting of the experimentally measured X-ray reflectivity data. 
DarpanX is implemented as a Python 3 module, and an API is provided to access the underlying algorithms. 
Here we present details of DarpanX implementation and its validation for different type multilayer structures. We also demonstrate the model fitting capability of DarpanX for experimental X-ray reflectivity measurements of single and multilayer samples.\\

\noindent \keywords{X-ray Astronomy, Instrumentation, X-ray optics, Multilayer mirrors}

\end{abstract}

\maketitle

\section{Introduction}
\label{sect:intro}  
X-ray reflection optics plays an integral part in the fields like X-ray 
astronomy, X-ray crystallography, X-ray microscopy, etc.  
At X-ray wavelengths, the refractive indices of all materials are close to unity, restricting the reflectivities to a very small grazing incidence angle.  
Hence, it is a common practice to employ small grazing incident angles 
to design X-ray reflecting systems.
Several optical designs, such as K-B optics\cite{ref-kb_optics}, Wolter 
type I, II, III\cite{wolter_1952}, etc. are developed for such 
grazing incidence applications.
The critical angle for total X-ray reflection is inversely 
proportional to the energy of the incident X-rays.
At energies higher 
than 10 keV, the critical angle becomes too small to design an efficient 
optical system.
Many applications of X-ray optics, particularly the astronomical X-ray 
optics, are restricted in energy range due to this limitation and hence
most astronomical X-ray telescopes in the past and present generation 
missions such as Einstein \cite{ref-heao2_mirrors}, ExOSAT \cite{ref-exosat_mirrors}, ROSAT \cite{ref-ROSAT_overview}, ASCA \cite{ref-ASCA_mirrors}, 
Chandra \cite{ref-chandra_overview},  XMM-Newton \cite{ref-xmm_overview}, Suzaku \cite{ref-suzaku_overview}, Swift \cite{ref-swift_overview}, AstroSat-SXT \cite{ref-astrosat_sxt} 
etc.
are limited to low energy (8 keV to 10 keV.) observations. 
There have been some attempts, such as the HERO balloon-born program \cite{ref-hero_mission}
to design X-ray telescopes with an extremely small incidence angle to
achieve focusing up to $\sim$60 keV. However, these have not to be culminated
in the space mission, mainly because of small effective area and requirement of very long focal length. 

Another approach to achieve reasonable X-ray reflectivity for incidence angles greater than the critical angle is to develop multilayer mirrors.
A multilayer mirror consists of a large number of alternate thin film layers 
of high-Z and low-Z materials deposited on a highly polished substrate.
The operational principle of multilayer mirrors is similar to Bragg's crystal. When X-rays are incident at an angle greater than the critical angle, 
a small fraction of the wave is reflected from the top layer, and
the rest of the wave is transmitted. The transmitted wave gets divided into transmitting and reflecting components at each layer interface.
The reflected components then get added up constructively, resulting in enhanced reflectivity. If the thickness of the alternative bilayers is
constant, the enhancement in reflectivity is usually limited to a
narrow energy band satisfying the Bragg condition. However, by
varying the thickness of the alternate high-Z and low-Z material 
in a controlled manner\cite{ref-joensen_porlaw,ref-supermirros_porlaw,ref-supermirrors1,ref-supermirros2}, it is possible to achieve broadband 
reflectivity for relatively larger incidence angles. With the advancement 
in the thin film fabrication technology, a variety of multilayer 
mirrors can be fabricated to reflect hard X-rays\cite{ref-wind_2003,ref-wind_2015}.
The NuStar hard X-ray telescope\cite{ref-Nustar_overview,ref-nustar_optics_overview} launched in 2012 is the  
X-ray astronomical telescope to employ such depth-graded multilayer mirrors. It used 
10 different recipes of multilayers of $W/Si$ and $Pt/C$\cite{ref-wsi_mirror_ex1} with the thickness varying from
2.50 nm to 12.8 nm to extend the higher energy cut-off till 79 keV. 
The Hitomi X-ray observatory\cite{ref-hitomi_overview} had a similar
hard X-ray telescope developed based on depth-graded multilayer 
mirrors\cite{ref-hitomi_supermirrors}.
Currently few proposals for a future hard X-ray focusing missions,
such as PolSTAR\cite{ref-polstar}, BEST \cite{ref-best}, FORCE \cite{ref-force},   
HEX-P \cite{ref-hexp}, 
InFOCuS~\citep{InFOCuS_2003}, X-Calibur~\cite{XCaliar_2014}\cite{X_Calibar_2020}, XL-Calibur~\cite{XL_Calibar_2021},  
etc.
that plan on using depth graded multilayer mirrors, are under various stages of consideration.
In India, we have initiated the development of hard X-ray optics for
a successor mission of AstroSat and have established
a multilayer mirror coating facility based 
on Magnetron Sputtering technique at Physical Research Laboratory (PRL), 
India.
In this context, we have developed a program ``DarpanX" (`Darpan' means mirror in Sanskrit) to aid the design of an X-ray 
multilayer mirrors as well as their characterization by using X-ray reflectivity (XRR) data. 

While the process for X-ray reflection from a multilayer mirror 
is governed by Bragg's law, the overall reflectivity of the mirror 
prominently depends on the surface micro-roughness, inter-layer 
roughness, interdiffusion at the interfaces, layer density, thickness, and uniformity of all layers. 
Precise knowledge of these parameters is essential for accurate modeling of X-ray reflectivity.
Traditional microscopic and optical interferometric techniques give information about the top surface roughness and uniformity. Hence, they cannot be used to model the mirror properties and performance completely. Transmission electron microscopy (TEM) gives information about the interlayer roughness and thickness, but it is not useful to probe the surface uniformity. 
TEM is also a destructive technique; hence it is not suitable to test the final product. 
XRR is a non-destructive technique that provides
information about the thickness of all layers, interlayer roughness, and
diffusion, the density of thin films, etc. This relatively inexpensive technique can also be performed at multiple energies to understand
the energy-dependent physical properties of the X-ray mirrors.

In order to estimate the parameters of multilayer mirrors from XRR 
measurements, it is necessary to have an efficient algorithm that calculates the X-ray reflectivity as a function for a given set of parameters and geometry. There are a few software programs such as IMD\cite{ref5}, PPM
(Pythonic Program for Multilayers)\cite{ref-ppm,ref-wsi_mirror_spiga}, 
GenX\cite{ref-genx}, Motofit \cite{ref-motofit}, Reflex \cite{ref-reflex},
etc. that are available to perform this task.
Most of these are available as standalone programs 
with a graphical user interface. 
Particularly the IMD software  
is widely used in the design and characterization of 
X-ray mirrors for astronomical applications. 
Our main objective of developing DarpanX is to 
implement the X-ray reflectivity calculation as a model 
compatible with the standard X-ray astronomical fitting software
such as XSPEC~\cite{ref-xspec} and ISIS~\cite{ref-isis}. Particularly,
the XSPEC software is widely
used and has advanced fitting methods such as genetic algorithms
and Markov Chain Monte Carlo (MCMC) for finding global minima
and thus can be efficiently used to measure various parameters
of the multilayer mirrors with the DarpanX. 
Further, the time required for fitting can be reduced by exploiting the parallel processing capabilities of XSPEC or that of DarpanX.  
DarpanX can also be used as a stand-alone tool for the design of multilayer mirrors with the flexibility to design any type of multilayer structure.

In this article, we describe the details of algorithms, implementation, 
and
validation of DarpanX.
Section \ref{sec-valid} provides a brief description
of DarpanX algorithm and implementations. Section~\ref{valid_algorithms} provides the validation of algorithms by the comparison of the DarpanX results with IMD. Section \ref{sec-exp_valid} describes the
experimental validation of DarpanX by using XRR measurements of 
single-layer ($W$ and $Si$) and multilayer ($W/B_4C$) samples 
and Section \ref{sec-summary} summarizes the work presented here.

\section{DarpanX: Algorithm, Implementation}\label{sec-valid}

\begin{figure}
\centering
  \centering
  \includegraphics[width=1.0\linewidth]{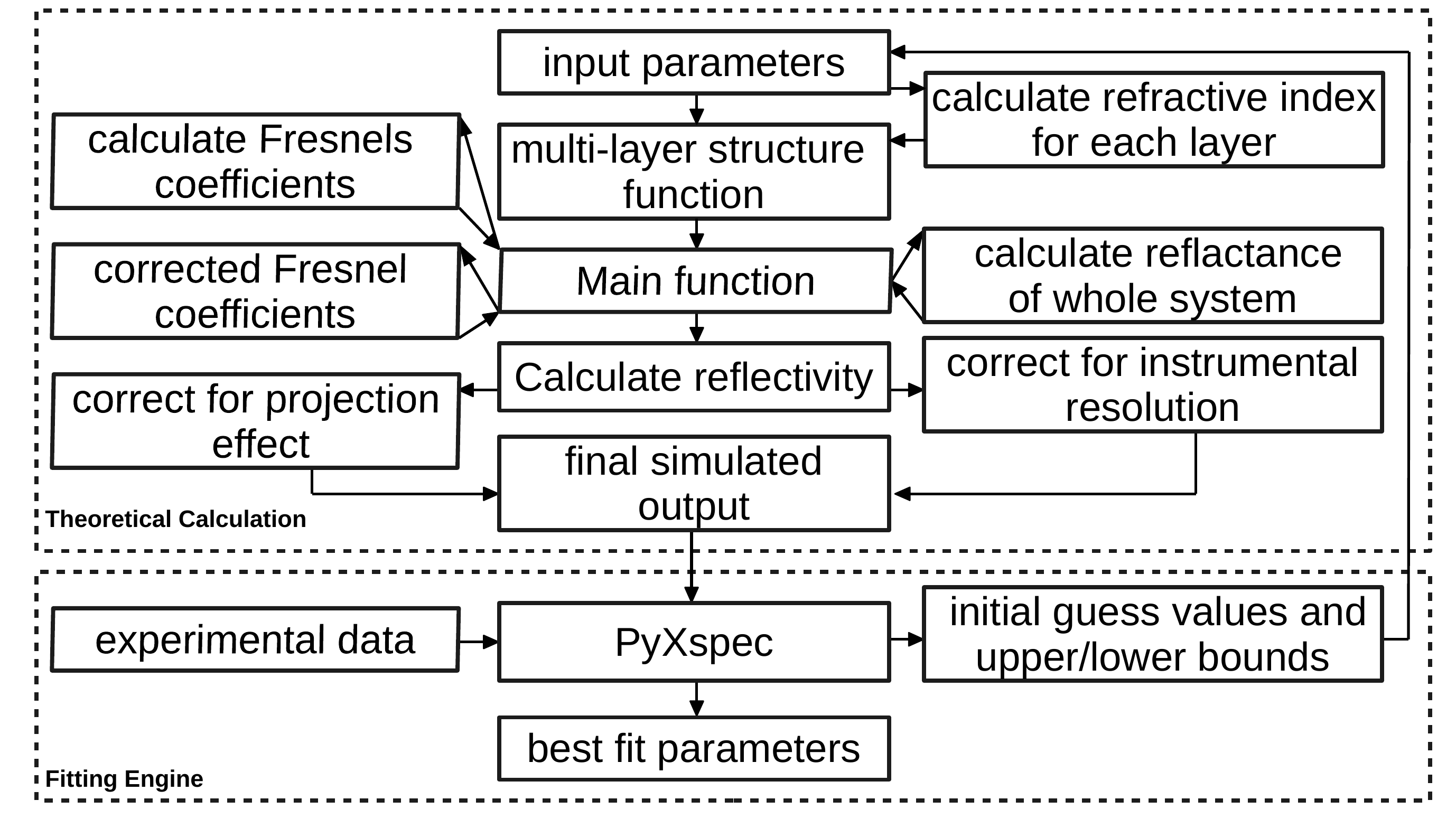}
  \caption{ Flow chart representing the algorithm employed in DarpanX and fitting DarpanX model in PyXspec. }
  \label{code_description}
\end{figure}

DarpanX is a program designed to compute the reflectivity of multilayer 
mirrors as a function of energy and incidence angle. 
Calculation of reflectivity and 
transmitivity in DarpanX is based on the Fresnel equations,
 modified for the finite surface roughness. A short description of theoretical calculation used in DarpanX is given in appendix-\ref{sec-our_code}.
A flowchart of the algorithm employed in DarpanX is given in Figure~\ref{code_description}. 
Two sub-blocks in the flowchart correspond to the theoretical computation 
of reflectivity and fitting of experimentally measured reflectivity data 
with the theoretical model.

DarpanX takes parameters like the number of layers (N), materials, thickness ($d$), 
density ($\rho$), and surface roughness ($\sigma$) of layers as input to construct the multilayer 
structure. 
For the computation of refractive indices of the materials, DarpanX uses the 
X-ray form factors ($f_1$ and $f_2$) in the energy range 0.001 keV to 433 keV
provided by NIST (National Institute of Standards and Technology) database~\cite{ref-nist_f1f2_data} for 92 elements from Z=1 to Z=92. 
Refractive index is calculated as~\cite{spiga_thesis} 
\begin{equation}\label{eq-ff2nk}
n=1-\frac{N_a r_e}{2\pi A}\lambda^2 \rho (f_1+if_2)
\end{equation} 
where, $\rho$, $A$, $N_a$, $r_e$, $\lambda$ are `Mass density', `Atomic weight',
`Avogadro number', 
`electron radius' and `wavelength' respectively. Accordingly, the refractive index profile for the multilayer structure is obtained.  
Then, Fresnel's coefficient for each interface is computed and corrected for surface imperfections. 
Using these modified Fresnel coefficients, 
the optical function for the complete system is calculated in a recursive manner. 

In order to compare the theoretical model of reflectivity with the XRR measurements, 
further corrections for projection effect and instrument angular resolution are required.  
At very low angles of the incidence onto a small mirror sample, only 
a fraction of the incident beam is covered by the reflecting surface 
of the sample and reflectivity needs to be modified for this projection 
effect by multiplying it with a factor $f$ defined as:
\begin{equation}
f=\frac{A_ \mathrm {r}}{A_ \mathrm {i}}
\end{equation}
where $A_r$ is the sample area along the incident beam direction and $A_i=L/sin(\theta)$ is the projected area of the incident beam on the sample, with $L$ being the beam 
diameter and $\theta$ the incidence angle. It is to be noted that this 
correction is required only when $A_r \leq A_i$. This is included in the 
DarpanX model calculation.
In order to take into account the finite angular resolution of the XRR instrument, the reflectivity  computed from the model is then convolved with a Gaussian profile, whose FWHM represents the instrumental resolution.

DarpanX implements these algorithms to compute X-ray reflectivity in Python 3. It is written as an object oriented program and is packaged as a module that can be imported in user program.
The theoretical calculations in the upper block of the flowchart shown in the Figure \ref{code_description} are implemented by the various methods of a class called `$ML\_Structure$'. To use the DarpanX as a stand-alone package, a higher level class named as `$Multilayer$' is implemented. 
It includes different methods for various applications such as calculating the optical functions, plotting the final output of a calculation, etc.
The `$Multilayer$' class also provides the flexibility to  
design any multilayer structure, like single-layer, constant period bi-layer, depth-graded, cluster-graded, or  any random layer structure defined by the layer materials and thicknesses. 
The details of the DarpanX classes, their methods, and how to use them are described in the user manual\cite{darpanx_usermanual}.

DarpanX can also be used as a local model in PyXspec (Python interface to XSPEC), where the experimental XRR data can be loaded and fitted with the reflectivity model to determine the parameters of the multilayer structure. 
Note that, the default XSPEC minimization method is a modified Levenberg-Marquardt algorithm but it is also possible to use other algorithm such as Minuit2 migrad method and Minuit2 simplex method as mentioned in the XSPEC user manual\cite{xspec_manual}. The results in the present work used the modified Levenberg-Marquardt algorithm. 
The lower block of Figure \ref{code_description} shows the fitting of experimentally measured reflectivity data  with the theoretical model by using PyXspec. 
A Python routine is included to link the DarpanX and PyXspec, which 
creates an object of the PyXspec class `$Model$'. This object can be used as a local model of PyXspec for the DarpanX module and can be accessed from the interactive Python shell.
Once DarpanX is loaded as a model in PyXspec, various parameters of the 
multilayer structure can be constrained to be constant or free for fitting. 
For free parameters, initial guess values and bounds are to be provided. 
The initial model function is computed with these values as input, and then observed XRR data can be fitted to obtain best-fit parameters.

Computation of refractive indices of various materials from their X-ray form factors (f1 and f2) are carried out by the methods of a class named `$nkcal$'. 
The $f_1$-$f_2$ data sets from NIST are reformatted and included with DarpanX. 
There is a provision to compute the refractive indices of compound if their density is provided as an input.
The optical constants for all the 92 elements 
as well as some common compound materials are pre-calculated using the `$nkcal$' class and are distributed with the DarpanX package. 
It is also possible for users to provide alternate data sets of form factors or optical constants in prescribed format (as described in the user manual\cite{darpanx_usermanual}) for use with DarpanX.
 
DarpanX software package also has provision to parallelize the computation of reflectivity over an array of energies or angles. This significantly reduces the time required to calculate optical functions for a multilayer structure with a large number of layers. For this purpose, it uses the Python multiprocessing library. Users can provide the number of cores to be used as an input to avail the parallelization option in DarpanX. Apart from this in-built parallelization in DarpanX, PyXspec has its own parallelization capability, where it performs multiple iterations required for fitting on different threads. Thus, for experimental XRR data fitting, one can use either the parallelization option in DarpanX or PyXspec.

DarpanX is distributed under the GNU General Public License (GPL).
It can run on any platform that supports Python. However, for the XRR data fitting, the platform should also support PyXspec.
DarpanX package is available on GitHub~\footnote{\href{https://github.com/biswajitmb/DarpanX.git}{https://github.com/biswajitmb/DarpanX.git}}.

\section{Validation of algorithms}\label{valid_algorithms}

In order to validate the DarpanX algorithms,
we compare the computed reflectivity against the results from the widely used software IMD.
The reflectivity calculation depends on the optical constants (refractive indices) database. DarpanX uses the database formed by using the X-ray form factors imported from NIST as described in Section~\ref{sec-valid}. 
IMD uses different databases by using the combined X-ray form factors from the Center for X-ray Optics (CXRO) and the Lawrence Livermore National Laboratory (LLNL) \cite{imd_manual}. 
Hence, for comparison of the reflectivity calculation from DarpanX with IMD, the reflectivity is calculated by DarpanX by using both the DarpanX database (database-1)  and the IMD database (database-2).
We have carried out the validation of DarpanX against IMD for an extensive
range of parameters and materials. Reflectivity is computed as a function 
of incident photon angles and energies with DarpanX for single layers of 
$Pt$, $Ni$, $C$, $W$, $Si$, $Cu$, $B_4C$, $Au$, $Ag$ etc. and multilayers of 
$W/Si$, $W/B_4C$, $Pt/C$, $Pt/SiC$, $Ni/C$, $Cu/Si$ etc. with different 
values of thickness/period ranging from 1$\textup{\AA}$-1000$\textup{\AA}$. 
A similar calculation is performed with IMD, and the results are compared. 
We find that the reflectivity computed from both software matches very 
well in all cases, if the same optical constant database is used. A few representative results are discussed below. 

\subsection{Single layer}
We consider the reflectivity for a single pure $Pt$ layer of thickness $100\textup{\AA}$ with an ideal interface. The calculation has been done for the reflectivity measurement as a function of the incident beam angle from the interface. The energy of the incident beam is 10.0 keV and 40.0 keV. In Figure \ref{fig3}, the outputs of both IMD (star and plus symbols) and DarpanX (solid and dashed lines) are shown. 
It can be observed that the calculation by DarpanX (solid lines in Figure~\ref{fig3}) exactly matches with the IMD result (star and plus symbols in Figure~\ref{fig3}) for both the energies when the IMD database (database-2) is used by DarpanX.
The DarpanX calculation by using the database-1 (dashed lines in Figure~\ref{fig3})  slightly differs from the IMD calculation, however, this is expected as the two databases (database-1 and database-2) are generated by using the X-ray form factors available from different sources. 
The dependence of critical angle ($\theta_c$) with the energy ($E$) of the incident X-ray beam on the reflecting surface, $\theta_c \propto \frac{1}{E}$ is clearly visible in  the Figure \ref{fig3}. With the increase of incident X-ray energy from 10 keV to 40 keV the critical angle ($\theta_c$) decreases from $0.46^{\circ}$ to $0.16^{\circ}$.
  
\subsection{Constant period Multilayer}
\begin{figure*}
\centering
\begin{subfigure}{.5\textwidth}
  \centering
  \includegraphics[width=1.0\linewidth]{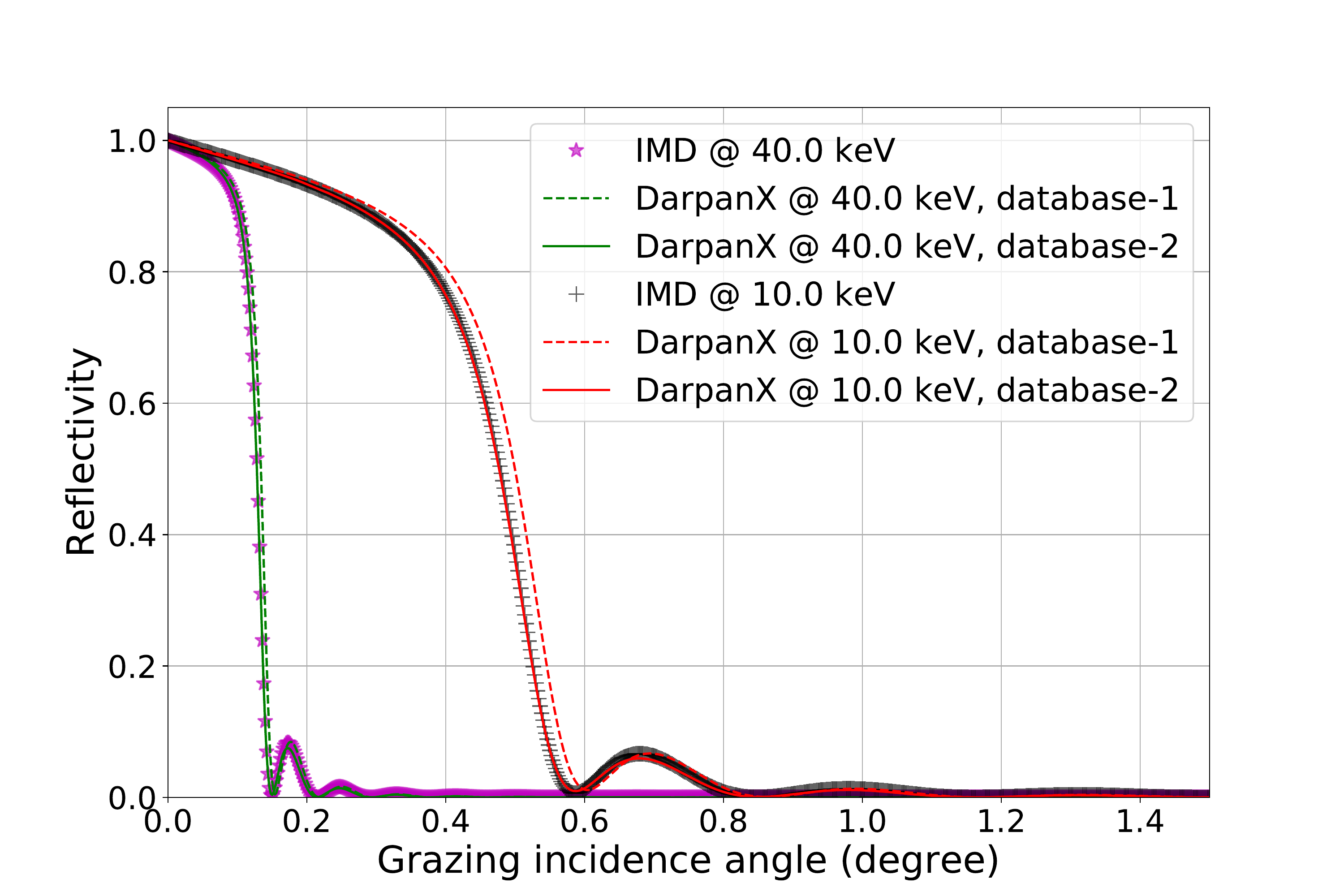}
       \caption{}
  \label{fig3}
\end{subfigure}%
\begin{subfigure}{.5\textwidth}
  \centering 
   \includegraphics[width=1.0\linewidth]{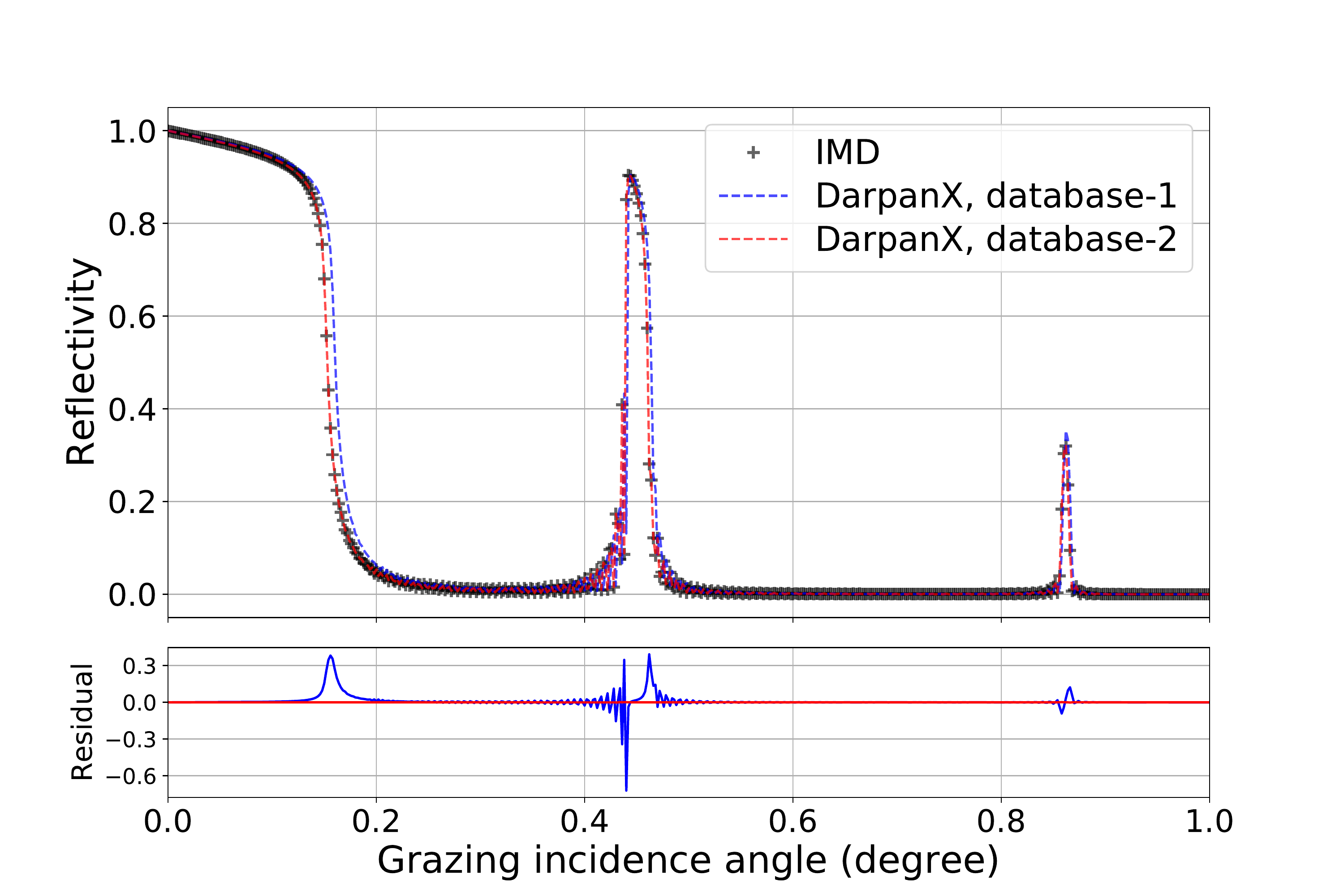}
  \caption{}
  \label{fig-NiC}
\end{subfigure}
 \caption{Comparison of reflectivity computed with DarpanX and IMD: (a) For single $Pt$ layer of thickness 100 $\textup{\AA}$ as a function of incident angle (b) For a  $Ni/C$ multilayer system with 60 bilayers of constant period of $50.0 \textup{\AA}$ and $\Gamma = 0.4$ at incident energy of 16.75 keV ($\lambda=0.74\textup{\AA}$). Here database-1 or database-2 represents, which optical constant database is used in the calculation. The database-1 is the DarpanX database formed by the X-ray form factors available from NIST and database-2 is that given in the IMD package formed by the combined X-ray form factors available from CXRO and LLNL.}
\end{figure*}

Constant period multilayers contain alternate layers of high Z and low Z materials of constant period (total thickness of high Z and low Z material of each bilayers). For verification of DarpanX, we calculate the reflectivity for a constant period of 60 bilayer $Ni/C$ system. 
We take the period of the system as $50.0\textup{\AA}$ with the gamma($\Gamma$) value (ratio of high-Z thickness to the total thickness of each bilayers) as 0.4.
Figure \ref{fig-NiC} shows the reflectivity of this multilayer system as a function of the incident angle at an energy of 16.75 keV.
From Figure \ref{fig-NiC}, it is clear that the reflectivity computed by DarpanX (red dashed) by using IMD database (database-2) at each Bragg peak, as well as below critical angle matches with the output of IMD (black plus symbols).
The blue dashed lines in the top panel of Figure~\ref{fig-NiC} shows the DarpanX calculation by using the database-1, where it slightly differs from the IMD calculation (black plus symbols).
The bottom panel shows residuals (between DarpanX and IMD calculations), which are negligible when the DarpanX used the database-2 (red solid line).
Finite residuals are present when the DarpanX used the database-1 (blue solid line).

\subsection{Depth-graded Multilayer}
\begin{figure*}
\centering
\begin{subfigure}{.5\textwidth}
  \centering
  \includegraphics[width=1.0\linewidth]{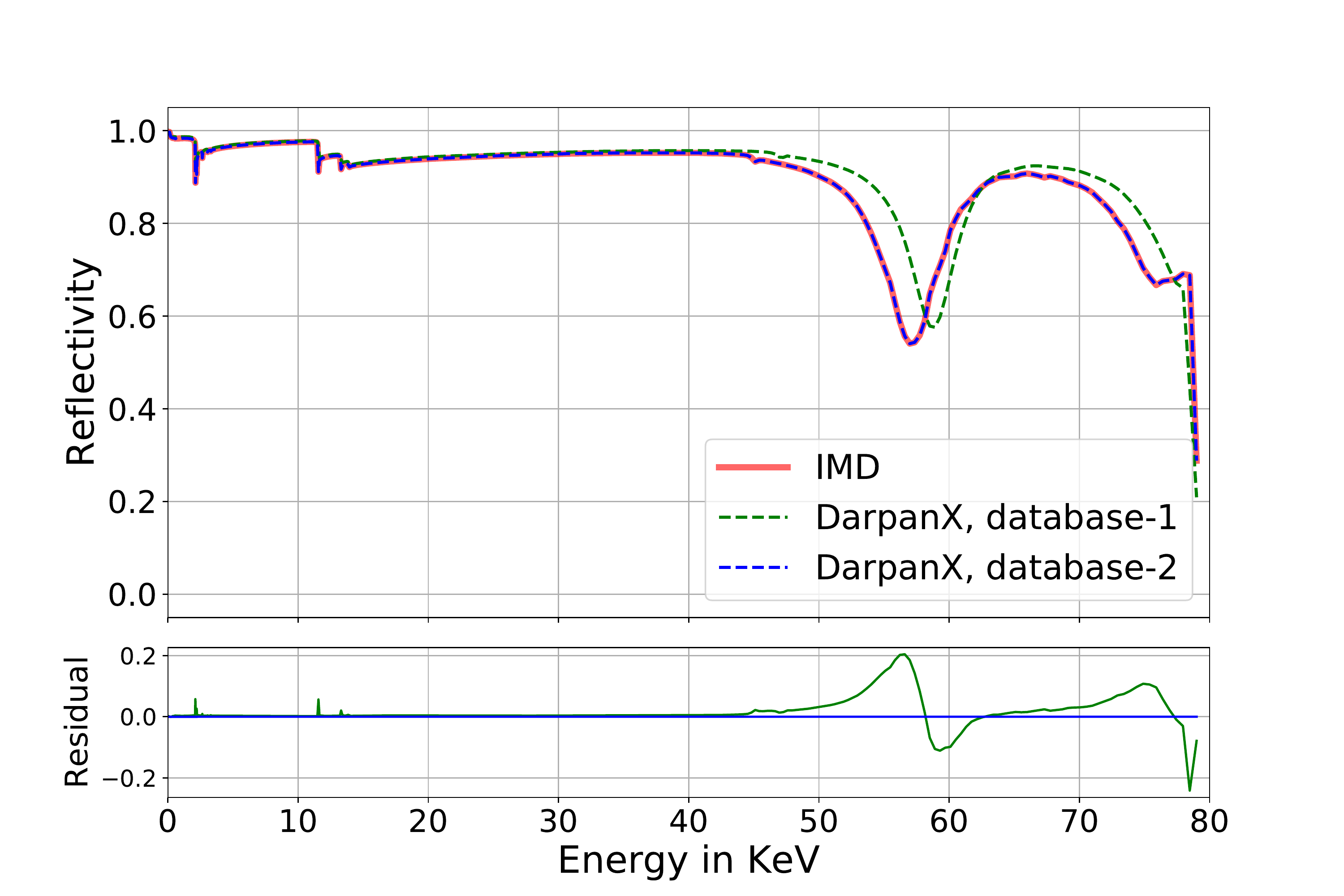}
       \caption{}
  \label{fig5}
\end{subfigure}%
\begin{subfigure}{.5\textwidth}
  \centering 
  \includegraphics[width=1.0\linewidth]{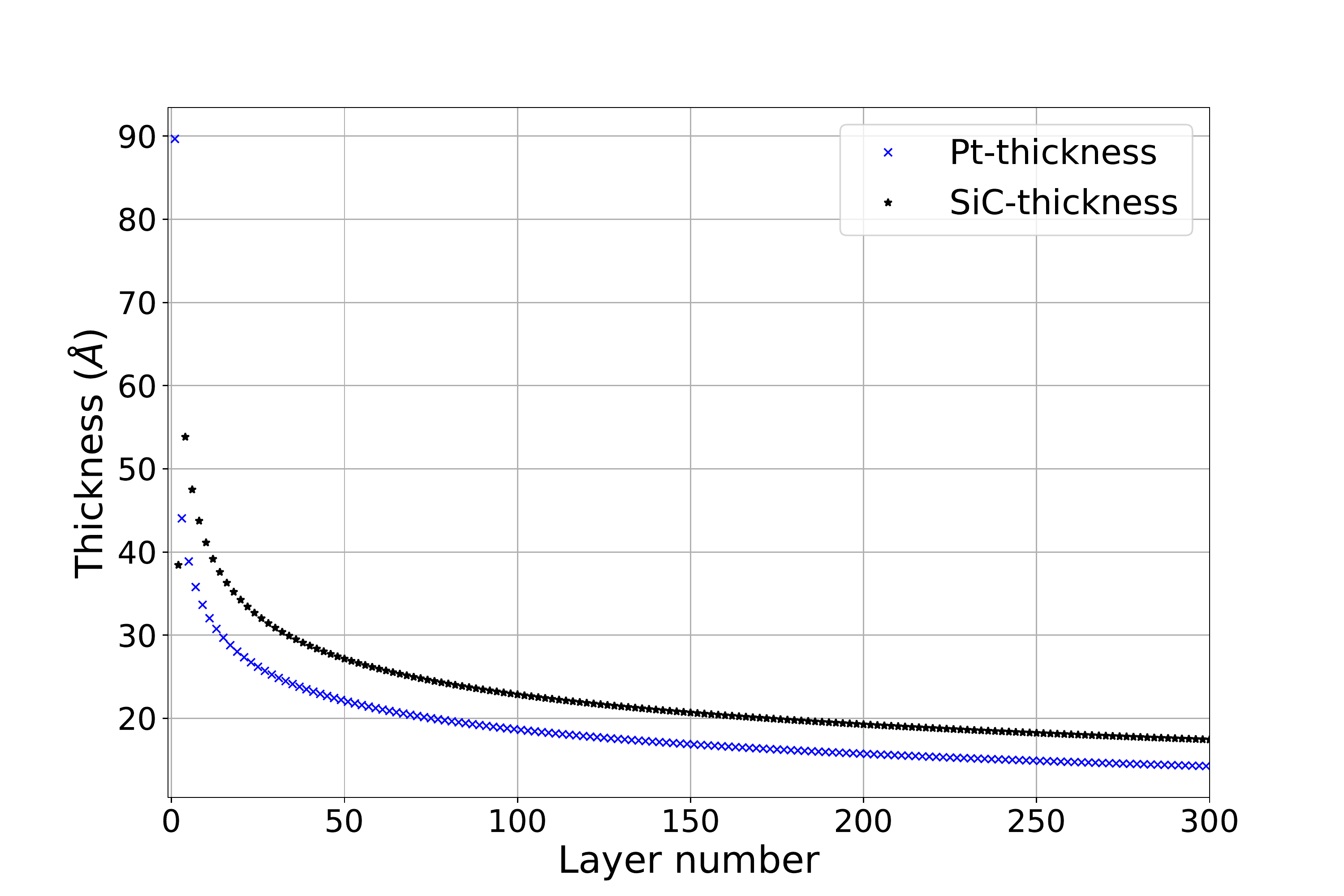}
  \caption{}
  \label{fig6}
\end{subfigure}
 \caption{{(a)Comparison of relefectivity of $Pt/SiC$ depth-graded system computed with DarpanX and IMD. Note that the higher value of $\Gamma_{top}$ as given in  Table \ref{pt-c_table}, applies only to the top bilayer (i.e. $d_{N=1}$ which has the period $d_{max}$ ). The thicker heavy material on the top improves total external reflectivity efficiency significantly below the critical energy, $E_c$ . Here database-1 or database-2 represents, which optical constant database is used in the calculation.
 (b) Thickness variation of the $Pt/SiC$ depth-graded system. }}
\end{figure*}

We calculate the reflectivity for a multilayer depth-graded $Pt/SiC$ system (Figure \ref{fig5}) of 150 bilayers, which is one of the NuSTAR multilayer recipe \cite{nustar_rec}. Here the period of the $Pt/SiC$ bilayer is changing from the top layer to the bottom layer according to the equation:\ref{eq-period_var} \cite{ref-joensen_porlaw,nustar_rec}.
\begin{equation}\label{eq-period_var}
d_i=\frac{a}{(b+i)^c} \hspace{1.5cm} 
\end{equation}
where, i=1,2...N is the number of bilayer, $a=d_{min}(b+N)^c$ ,  $b=\frac{1-N.k}{k-1} \hspace{0.5cm}$  and  $\hspace{0.2cm} k=(\frac{d_{min}}{d_{max}})^{\frac{1}{c}}$. Here $d_{min}, \hspace{0.2cm} d_{max}$ are the bottom most and top most d-spacing (or period) respectively and $c$ controls the slop between these extreme value over the N-bilayers. 
Figure \ref{fig6} shows the thickness of each layer.
Table-\ref{pt-c_table} gives the parameters of the depth-graded multilayer.

\begin{table*}
\begin{center}
\begin{tabular}{|r|c|c|c|c|c|c|c|l|} 
\hline
  $Material$  & $d_{max}$ $({\textup{\AA}})$ & $d_{min}$ $(\textup{\AA})$ & $N$ & $\Gamma_{top}$ & $\Gamma$ & $c$ & $\sigma$ $(\textup{\AA})$ \\

  \hline  
    $Pt/SiC$ & 128.10 & 31.70 & 150 & 0.7 & 0.45 & 0.245 & 4.5\\

  \hline 
\end{tabular}
\end{center}
\caption{Multilayer depth-graded recipe details. $d_{max}$ and $d_{min}$ is the maximum and minimum period corresponding to the top and bottom most bilayer. $N$ and $\Gamma$ are the number of bilayers and gamma factor respectively. $\Gamma_{top}$ is the gamma factor corresponding to the top most layer. $c$ controls the slope between these extreme values of $d_{max}$ and $d_{min}$ over the N layers. }
\label{pt-c_table}
\end{table*}

\noindent
The reflectivity has been calculated as a function of incident energy varying from 0.01 keV to 79.0 keV with the incident angle of the beam kept at a constant value of 0.077$^o$.
Figure \ref{fig5} shows the reflectivity computed from DarpanX and IMD, which match very well, if the same optical constant database is used.
The bottom panel shows residuals (between DarpanX and IMD calculations), which are negligible (blue solid line) when both the DarpanX and IMD uses the same database (database-2). Finite residuals (green solid line) are present when the DarpanX used the database-1.

The accuracy of the DarpanX to calculate the X-ray reflectivity of single-layer or multilayer system is established with these results. 

\section{Experimental validation}\label{sec-exp_valid}
DarpanX model is further validated experimentally by employing it in the analysis of
X-ray reflectivity measurements of single and multilayer samples. Single-layer (W and Si) and multilayer ($W/B_4C$ with N=170 and $W/B_4C$ with N=50, where N is the number of bilayers)
thin film samples are prepared, and reflectivity measurements are carried out for these samples. The observed data is
then 
fitted with the DarpanX.   
\subsection{Samples and XRR measurements}\label{sec-ML_sample}
Single-layer samples of W and Si used in this work are prepared using the 
RF magnetron sputtering system in PRL. This system has two stationary 
targets placed at the center of a vacuum chamber, each having size of
$4\times30$ $\textrm{cm}^2$ (width $\times$ length). Substrates are placed on a rotating platform in front of the target at different distances. 
For the purpose of characterization of the sputtering system, single 
layers of each target material are deposited on a $20\times 20\times 0.3$ ${\textrm{mm}}^3$ (width $\times$ length $\times$ height) 
polished borosilicate SCHOTT glass substrate, with various operating parameters of the sputtering system. Two of these samples with a coating of 
Si and W are used in this study. 

X-ray reflectivity measurements for single-layer samples were carried out using 
High-Resolution X-ray Diffraction (HRXRD) system by Bruker (Bruker D8 DISCOVER) 
at Space Applications Center (SAC), Ahmedabad. XRR measurements were carried out at incident X-ray energy of 8.047 keV with the incident beam slit and detector slit of each 0.2 mm. XRR scan measurements for each sample were obtained after carrying out the alignment of the system.    

Two constant period $W/B_4C$ multilayer samples are also used for the experimental 
validation. This samples are deposited on a 0.5 mm thick n-type $Si$ substrate of 
dimension 30 $\times$ 30 $\textrm{mm}^2$.
They consists of 170 and 50 number of bilayers fabricated by using the facility at 
Raja Ramanna Center for Advanced Technology (RRCAT), Indore. 
XRR measurements for these samples were carried out with the facility at RRCAT.
The details of the sample and XRR measurements are given in Panini et
al. (2018)\cite{panini_2017}\cite{panini_2018}.

\subsection{XRR Analysis with DarpanX}\label{sec-XRR_analysis}

\subsubsection{Single-layer}\label{sec-sl_analysis}

\begin{figure*}
\centering
\begin{subfigure}{.5\textwidth}
  \centering
  \includegraphics[width=1.0\linewidth]{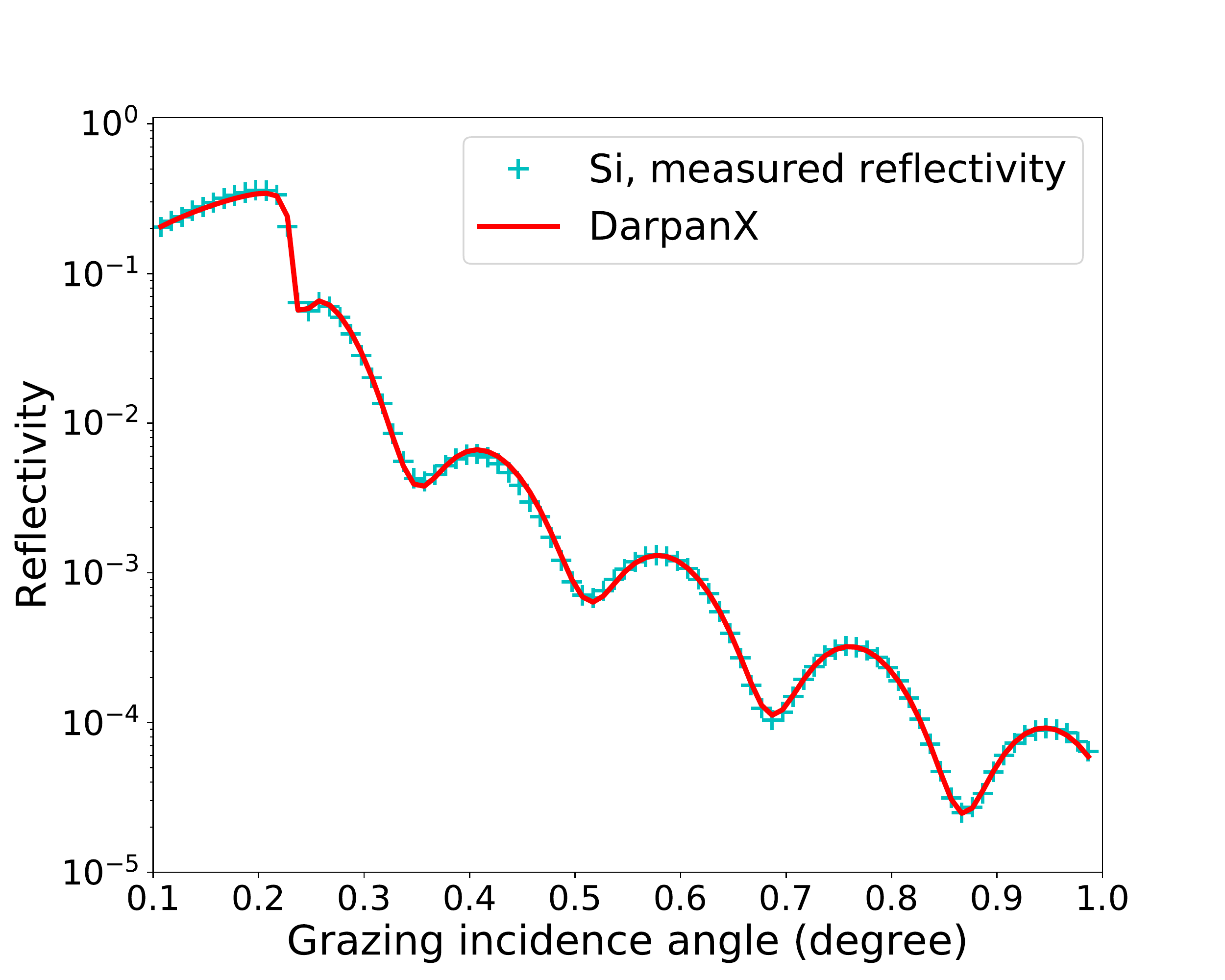}
       \caption{}
  \label{fig:fit_pyxspec_Si}
\end{subfigure}%
\begin{subfigure}{.5\textwidth}
  \centering 
  \includegraphics[width=1.0\linewidth]{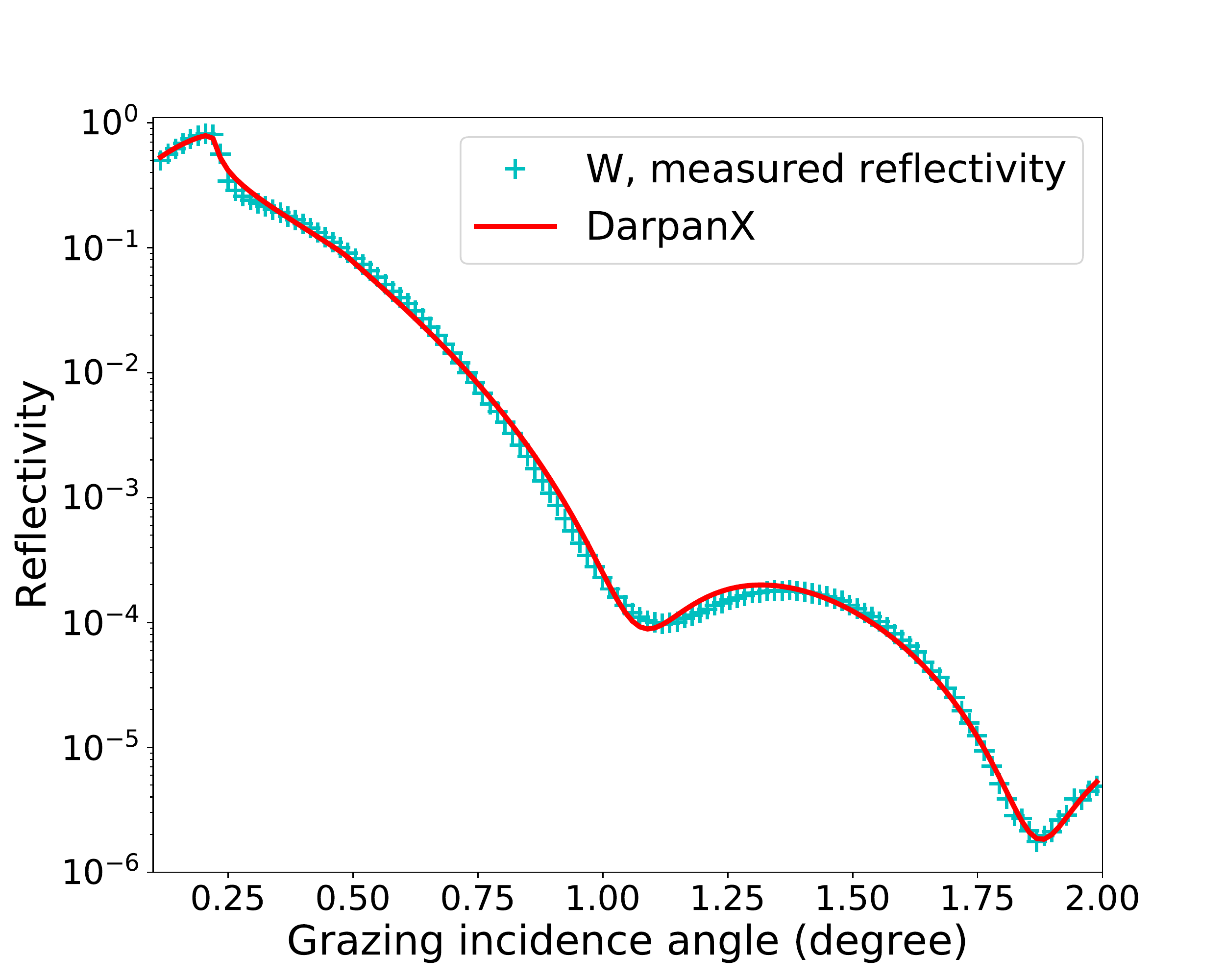}
  \caption{}
  \label{fig:fit_pyxspec_Si}
\end{subfigure}
 \caption{ XRR reflectivity data(cyan) over plotted with DarpanX model (red) of (a) Single Si-layer and (b) single-W layers deposited at PRL coating facility. The fitted parameters are summarized in Table~\ref{W_and_Si_m_table}.}
\label{fig-fit_pyxspec_single_layers}
\end{figure*}

XRR measurements (cyan plus points) for $Si$ and $W$ single layers are shown in Figure \ref{fig-fit_pyxspec_single_layers} left and right panels respectively.
As the top layer of the thin film exposed to the atmosphere would get oxidized, a thin oxide layer of $SiO_2$ (above $Si$) and $WO_3$ (above $W$) are included in the models used to fit the XRR data.
We keep both the thicknesses of oxide ($SiO_2$/$WO_3$) and material ($Si$/$W$) layers as free variables for the DarpanX fitting. 
Along with thicknesses, we consider, oxide and material densities, surface roughnesses ($\sigma$) of the three interfaces ($ambient/oxide$ , $oxide/material$ and $material/substrate$) as free parameters. 
The instrumental resolution defined as the standard deviation of the Gaussian convolution function is also considered as a free parameter.
Fit results from 
DarpanX (red) are overplotted with the XRR data in Figure \ref{fig-fit_pyxspec_single_layers}.
The best-fit parameters obtained 
are given in Table~\ref{W_and_Si_m_table}. 
Note that the $W$ density for the $WO_3/W$ sample is very low (close to the density of its oxide) compared to the original density. 
This may be because the thickness of the $W$ layer is very less. As a result, most of it oxidized.

\begin{table*}
\begin{center}
\resizebox{17 cm}{!}{
\begin{tabular}{|l|c|c|c|c|c|c|c|c|c|c|r|} 
\hline
\large$Material$ &\Large$d_{layer}$ & \Large$d_{oxide}$ & \Large$\rho_{oxide}$ &\Large$\rho_{layer}$&\Large$\sigma_1$ & \Large$\sigma_2$ &\Large$\sigma_3$&\large$Resolution$ \\
 &$(\textup{\AA})$ &  $(\textup{\AA})$ & ($\textrm{g/cm}^3$) & ($\textrm{g/cm}^3$) & $(\textup{\AA})$ & $(\textup{\AA})$ &  $(\textup{\AA})$ &  (deg) \\[1ex]
  \hline 
      
      $SiO2/Si$&$161.26$ & $71.65$ & $2.17$ & $2.10$ & $9.52$ & $1.10$ & $3.13$ & $0.001$ \\

      $WO_3/W$ &$32.31$ & $14.93$ & $2.72$ & $7.94$ & $4.11$ & $9.84$ & $4.34$ & $0.001$ \\
  \hline 
\end{tabular}
}
\end{center}
    \caption{Fitted parameters of the XRR measurement of the single layer samples. $d_{layer}$ and $d_{oxide}$ are the thicknesses of oxide and coating materials respectively. Same as thicknesses, $\rho_{oxide}$ and $\rho_{layer}$ are the densities of oxide and coating materials. $\sigma_1$ , $\sigma_2$ and $\sigma_3$ are the surface roughness of the three interfaces (ambient/oxide, oxide/layer, and layer/substrate) as discussed in the text.}
\label{W_and_Si_m_table}
\end{table*}

\subsubsection{Multilayer}

\begin{figure*}
\centering
\begin{subfigure}{.5\textwidth}
  \centering
   \includegraphics[width=1.0\linewidth]{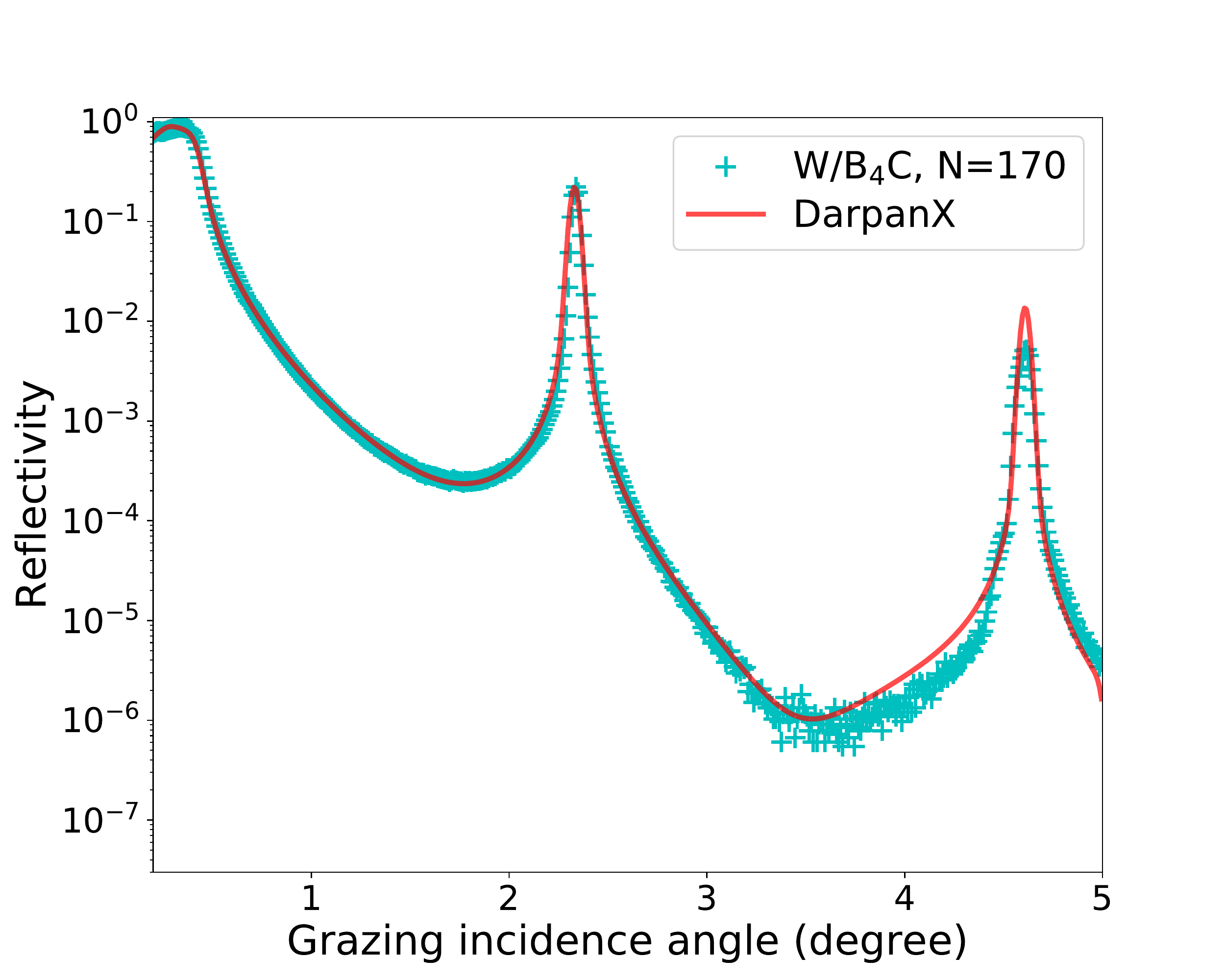}
  \caption{}
  \label{fig-panini_d19}
\end{subfigure}%
\begin{subfigure}{.5\textwidth}
  \centering 
  \includegraphics[width=1.02\linewidth]{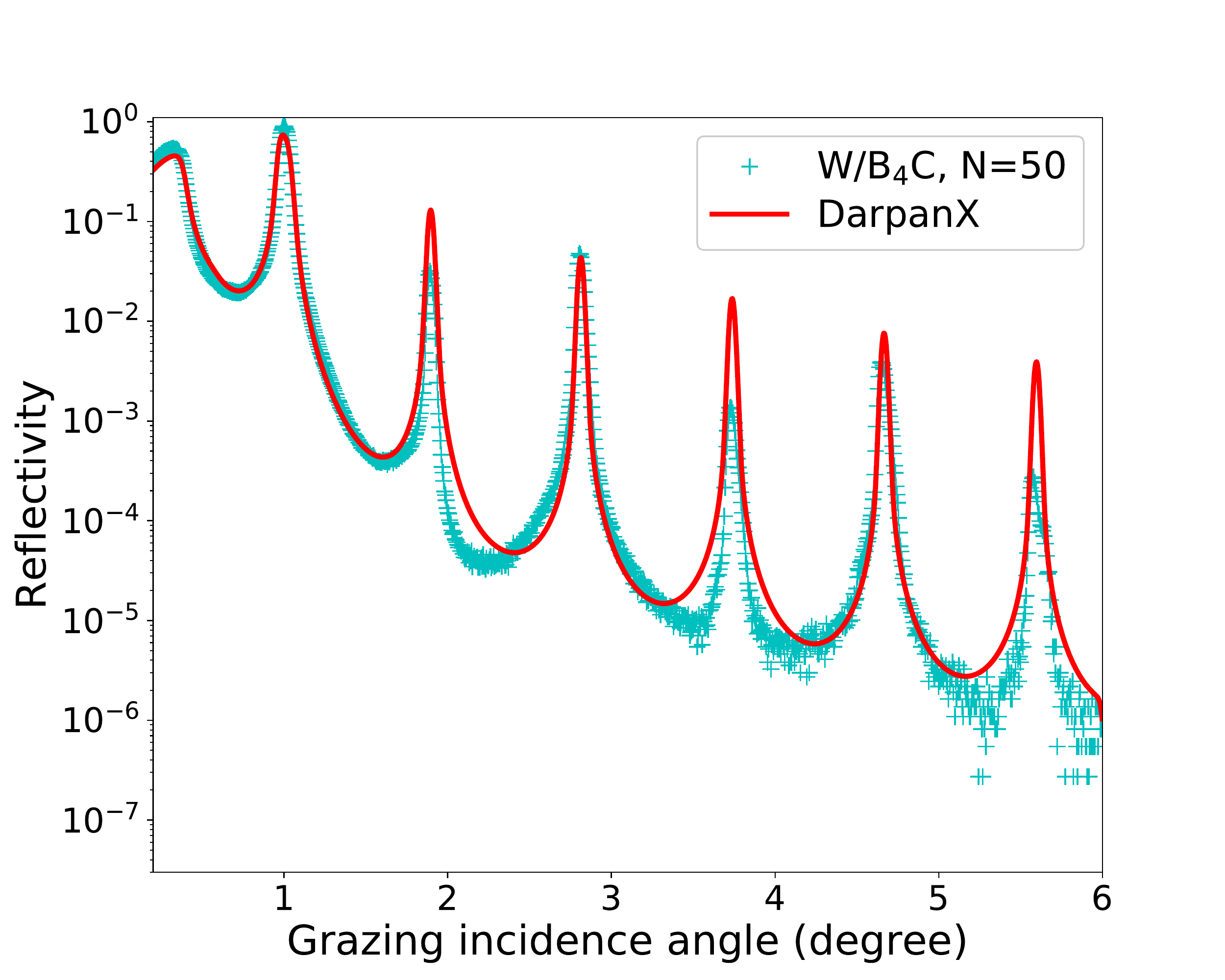}
  \caption{}
  \label{fig-panini_d44}
\end{subfigure}
 \caption{XRR reflectivity data(cyan) of $W/B_4C$ multilayer sample (with N number of bi-layers) fitted (red) with constant period bilayer model: (a) N=170, (b) N=50. The fitted parameters are summarized in the first and second row of Table~\ref{panini_samples_table}.}
\label{fig-ml_fit}
\end{figure*}

Figure \ref{fig-ml_fit} shows the XRR measurements (cyan plus points) of the multilayer samples of $W/B_4C$ (number of bilayers, N=170 and N=50).
In a first step data are fitted with DarpanX by considering a constant period bilayer models (number of bilayers, N=170 and N=50 respectively).
For the analysis, the multilayer period ($d$ = total thickness of $W$ and $B_4C$), gamma factor ( $\Gamma$ = ratio of $W$ thickness to the period), along with the surface roughnesses of the $ambient/W$ ($\sigma_1$), $W/B_4C$ ($\sigma_2$), $B_4C/W$ ($\sigma_3$) and substrate ($\sigma_4$) interfaces are kept as a free variable.
The best fit results are overplotted on the observation in Figure \ref{fig-panini_d19} and \ref{fig-panini_d44}, and the resultant parameters are summarised in the first and second row of Table~\ref{panini_samples_table}.

\begin{table*}
\begin{center}

\resizebox{17 cm}{!}{
\begin{tabular}{|r| c| c|c| c| c| c| c| c| c| c|l| } 
\hline
   \large$ Sample$ &\large$ Bilayers$ &\large $Model$ & {\Large$d$}  & {\Large$\Gamma$} & \Large {$\rho$}\small $_{ W}$  & \Large {$ \rho$}\small $_{ {B_4C}}$  &\Large $ \sigma_1 $ &\Large $ \sigma_2 $ &\Large $ \sigma_3$ &\Large$ \sigma_4$  & \large$ Resolution $ \\
  
  & ($N$) & & $(\textup{\AA})$  &   & ($\textrm{g/cm}^3$) & ($\textrm{g/cm}^3$) & $(\textup{\AA})$ &   $(\textup{\AA})$ &  $(\textup{\AA})$&$(\textup{\AA})$ &  ($\deg$) \\[1ex]

  \hline  
    $W/B_4C$ &170 & BL & 19.19 &   0.39 & 19.61 & 2.90 & 3.17 & 12.24 & 1.64 & 19.37 & 0.015 \\ [1ex]
    
    $W/B_4C$ & 50 &BL & 47.30  &  0.41 & 20.65 & 1.85 & 7.09 & 1.10 & 6.19 & 13.37 & 0.015 \\[1ex] 
    
    $W/B_4C$ & 170 & CG1 & -  &  - & 20.48 & 2.23 & 8.60 & 2.15 & 8.92 & 13.85 & 0.015 \\[1ex]
    $W/B_4C$ & 50 & CG1 & -  &  - & 17.16 & 2.34 & 2.66 & 1.66 & 1.64 & 13.37 & 0.015 \\[1ex]
     $W/B_4C$ & 170 & CG2 & -  &  - & 19.69 & 2.30 & 8.18 & - & - & 13.97 & 0.015 \\[1ex]
    $W/B_4C$ & 50 & CG2 & -  &  - & 16.09 & 2.25 & 2.64 & - & - & 13.20 & 0.015 \\[1ex]
  \hline 
\end{tabular}}
\end{center}
\caption{Fitted parameters of the Multilayer samples. {\large$d$} and {\large$\Gamma$} are  period and gamma factor. {\large$\rho$}$_{W}$ and {\large$\rho$}$_{B_4C}$ are the density of $W$ and $B_4C$ respectively. $\sigma_1$, $\sigma_2$, $\sigma_3$, and $\sigma_4$ are the surface roughness of $ambient/W$, $W/B_4C$, $B_4C/W$ interfaces and substrate respectively. Resolution is the instrumental angular resolution in degree. BL represent the Bi-Layer model. CG1 represents the Cluster-Graded model with different values of period and gamma at each block (Figure~\ref{fig-clustergradedfit_sigmacons_fitpar}). CG2 represent the model with different roughness values ($\sigma_2$ and $\sigma_3$), period, and gamma value  at each block (Figure~\ref{fig-clustergradedfit_sigmavarying_fitpar}).}
\label{panini_samples_table}
\end{table*}

\begin{figure*}
\centering
\begin{subfigure}{.5\textwidth}
  \centering
   \includegraphics[width=1.02\linewidth]{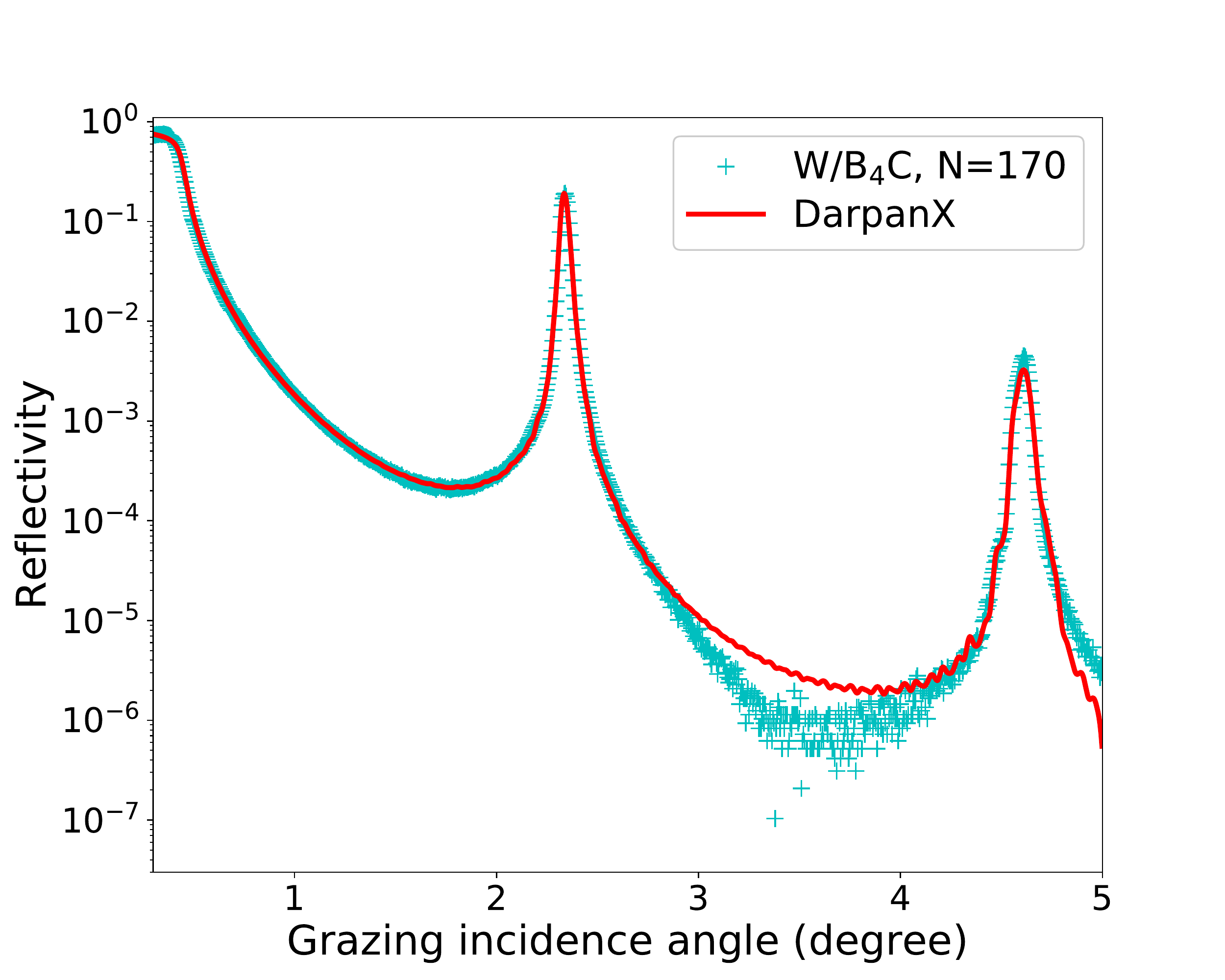}
  \caption{}
  \label{fig:clustergradedfitd19_sigmacons}
\end{subfigure}%
\begin{subfigure}{.5\textwidth}
  \centering 
  \includegraphics[width=1.02\linewidth]{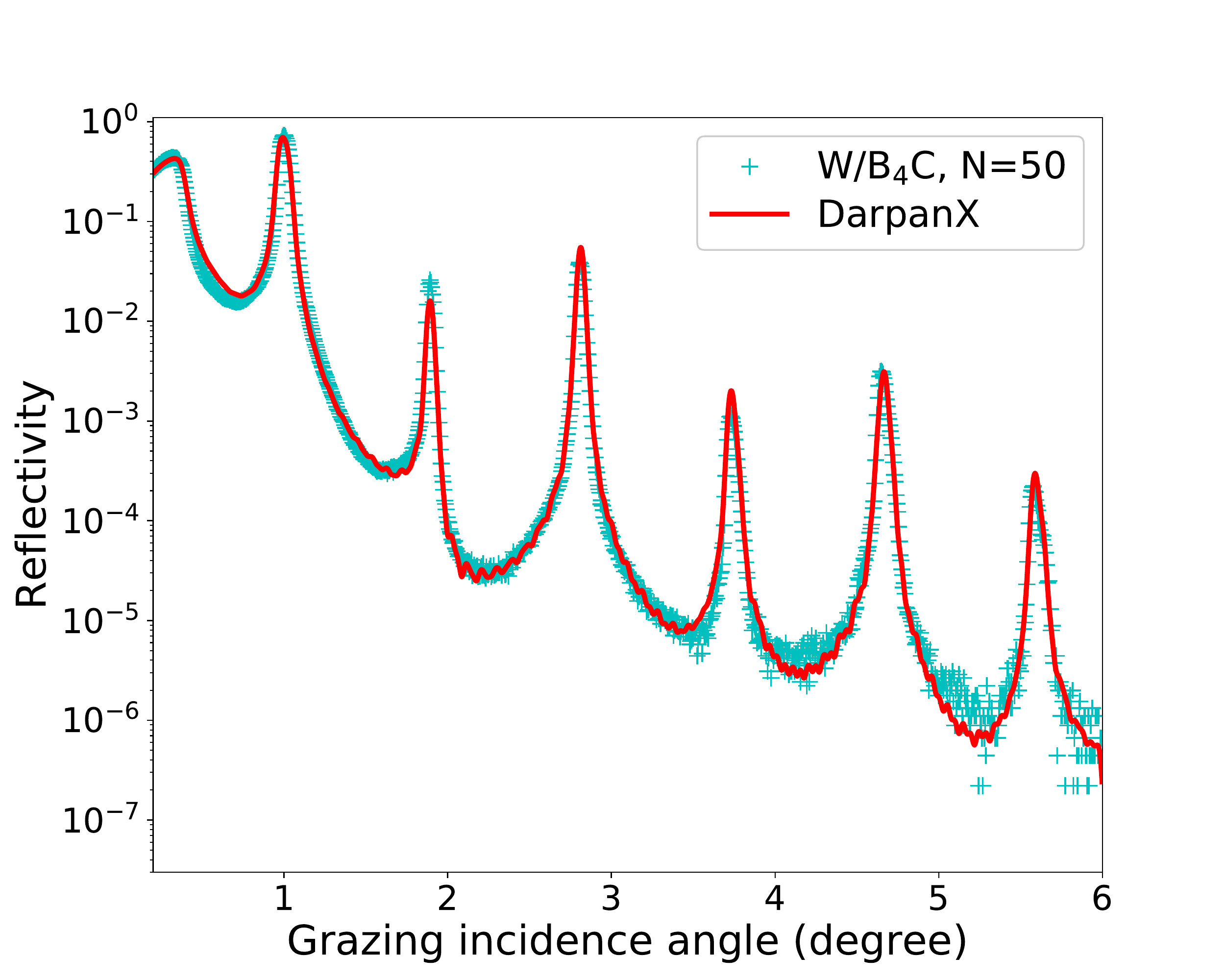}
  \caption{}
  \label{fig:clustergradedfitd44_sigmacons}
\end{subfigure}
 \caption{XRR reflectivity data(cyan) of $W/B_4C$ multilayer sample (with N number of bi-layers) fitted (red) with a cluster-graded model and considering the period and gamma are varying ( Figure~\ref{fig-clustergradedfit_sigmacons_fitpar} ) in stacks. (a) N=170, (b) N=50.}
\label{fig-clustergradedfit_sigmacons}
\end{figure*}

\begin{figure*}
\centering
\begin{subfigure}{.48\textwidth}
  \centering
   \includegraphics[width=1.01\linewidth]{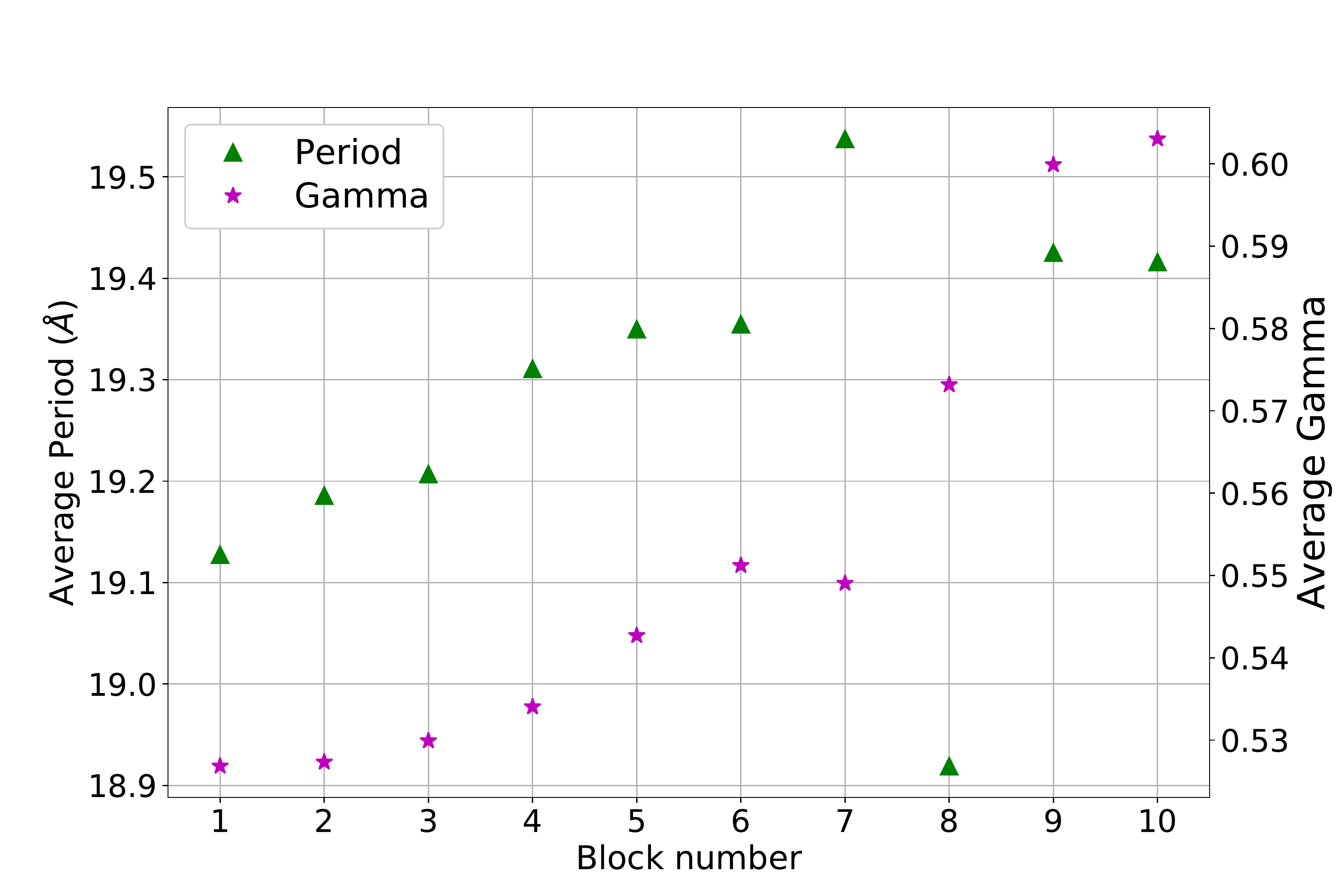}
  \caption{}
  \label{fig:clustergradedfitd19_sigmacons_fitpar}
\end{subfigure}%
\begin{subfigure}{.5\textwidth}
  \centering 
 \includegraphics[width=1.02\linewidth]{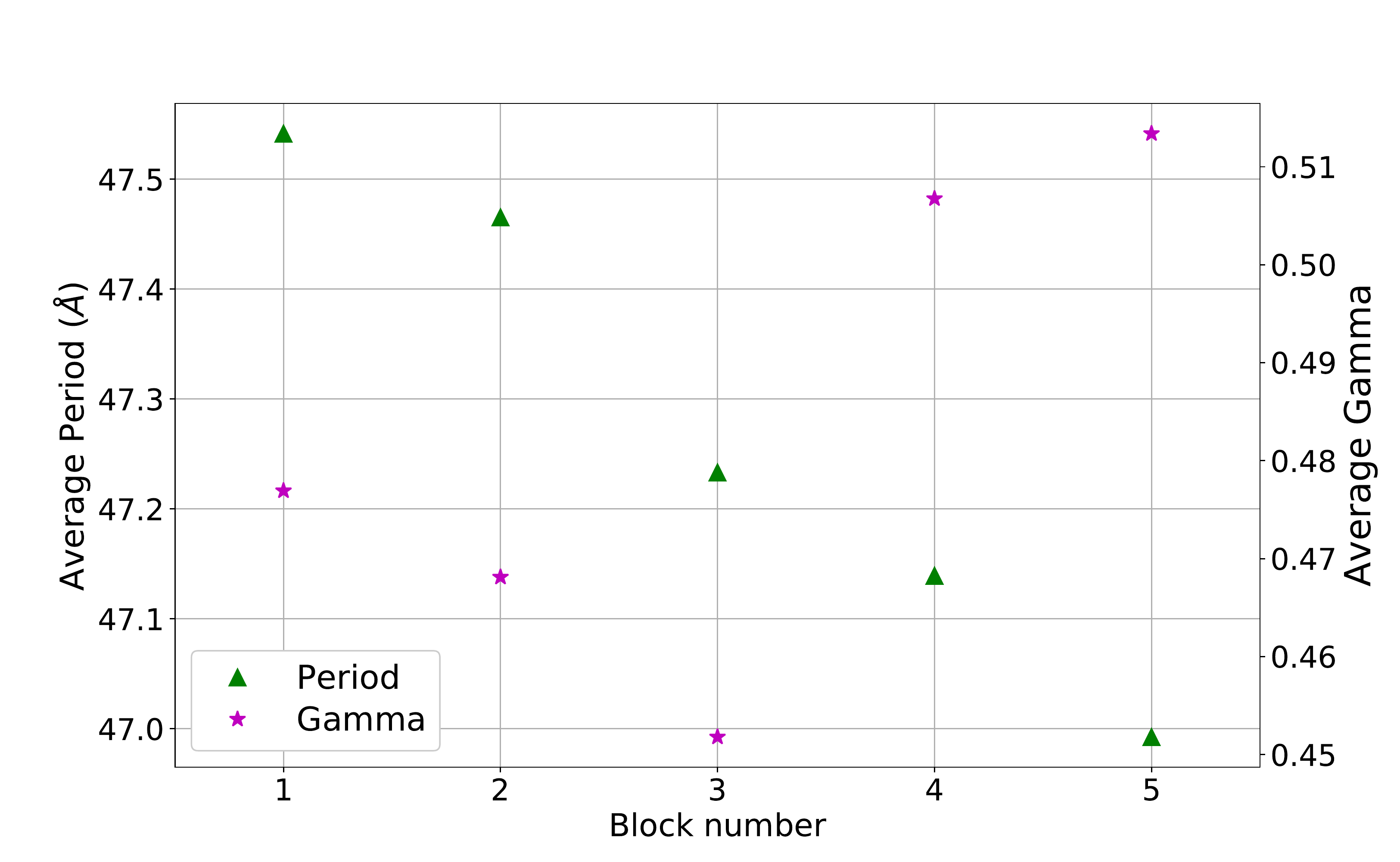}
  \caption{}
  \label{fig:clustergradedfitd44_sigmacons_fitpar}
\end{subfigure}
 \caption{Variation of average $period$ and $\Gamma$ of $W/B_4C$ bilayers at each block of cluster-graded model for samples: (a) N=170 and (b) N=50.}
\label{fig-clustergradedfit_sigmacons_fitpar}
\end{figure*}

The best fit model of the experimental XRR scan result, as shown in Figure \ref{fig-panini_d19} and \ref{fig-panini_d44}, indicates that there is a deviation between the experimental data and fitted model at the higher order Bragg peaks. 
An additional peak (at $\theta \approx 4.5$) in Figure \ref{fig-panini_d19} just below the second Bragg peak suggests that there might be variation in thickness within the multilayer system, i.e., it deviates slightly from the constant period. 
If we consider the parameters of each layer, such as thickness, density, and surface roughness, to be independent, there would be large number of free parameters for fitting.
To avoid this, we segregated the N=170 and N=50 bilayer systems as a succession of 10 and 5 blocks formed by 17 and 10 bilayers each.
This type of multilayer structure is called cluster graded\cite{ref-joensen_porlaw}. 
As a result, any period variation of the layers will be approximately modeled. 
We assume the density of $B_4C$, $W$, and the surface roughnesses of $B_4C/W$, $W/B_4C$- interfaces remain the same over the multilayer stack and keep them as free variables for fitting. 
The multilayer XRR data is fitted with an overall set of 27 (for N=170) and 18 (for N=50) free parameters, and the result is shown in Figure \ref{fig:clustergradedfitd19_sigmacons} and \ref{fig:clustergradedfitd44_sigmacons}. Figure \ref{fig-clustergradedfit_sigmacons_fitpar} shows the variation of the $period$ and $gamma$ to each block of 17 and 10 bilayers through the whole multilayer stacks.
The best-fit parameters that are common to all blocks are shown in the third (for N=170) and fourth (for N=50) row of Table~\ref{panini_samples_table}.

\begin{figure*}
\centering
\begin{subfigure}{.5\textwidth}
  \centering
   \includegraphics[width=1.02\linewidth]{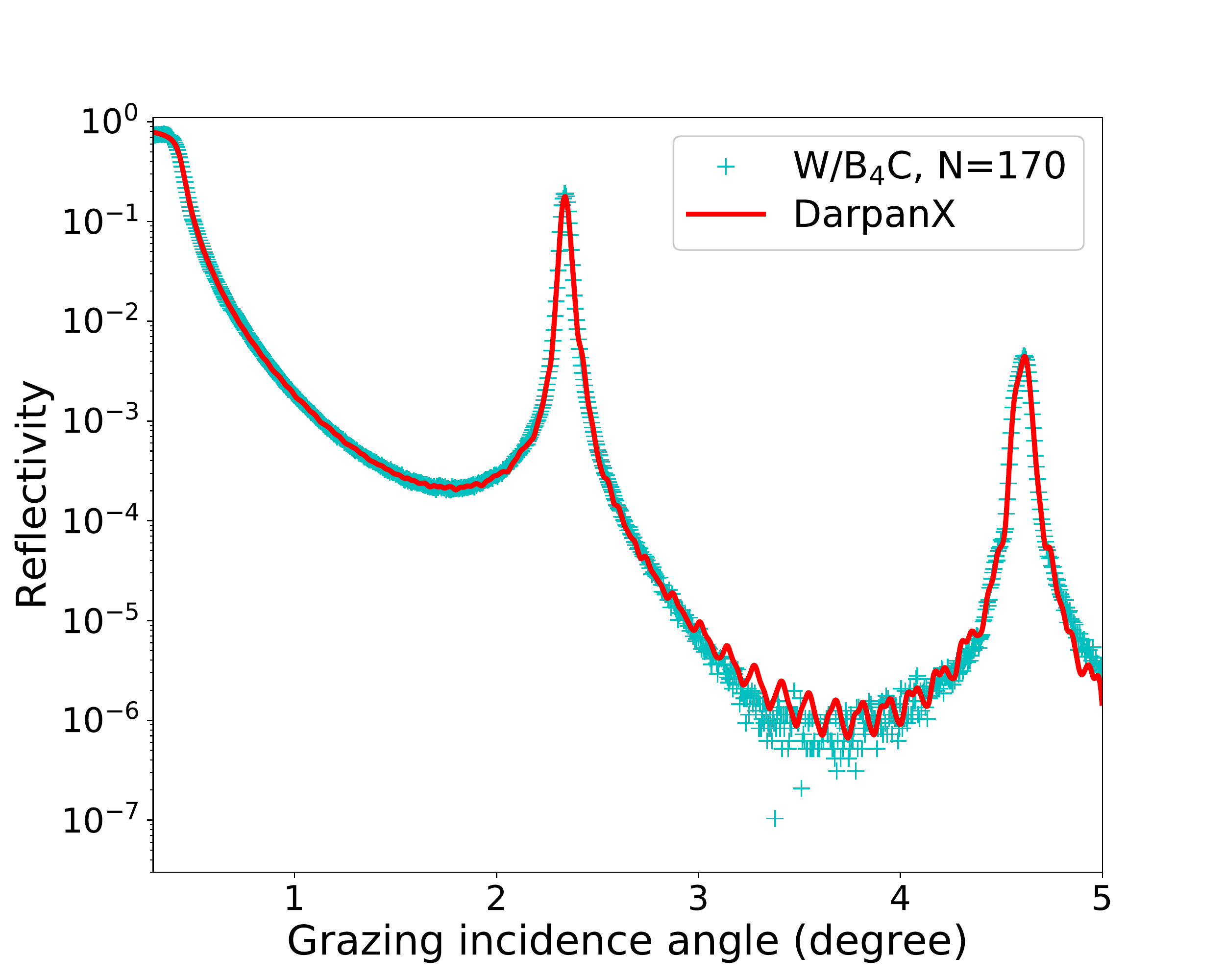}
  \caption{}
  \label{fig:clustergradedfitd19_sigmavarying}
\end{subfigure}%
\begin{subfigure}{.5\textwidth}
  \centering 
  \includegraphics[width=1.02\linewidth]{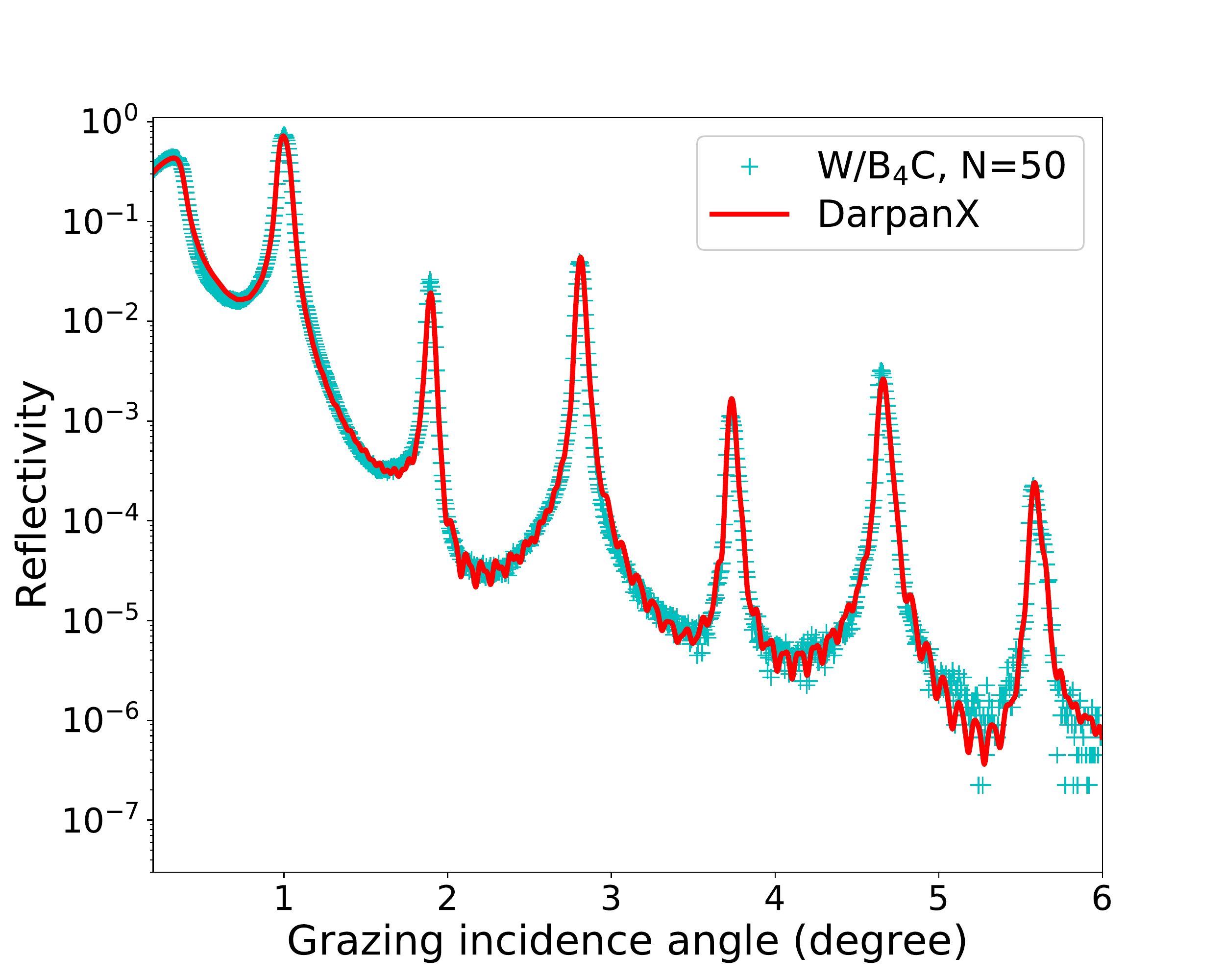}
  \caption{}
  \label{fig:clustergradedfitd44_sigmavarying}
\end{subfigure}
 \caption{XRR reflectivity data(cyan) of $W/B_4C$ multilayer sample (with N number of bi-layers) fitted (red) with a cluster-graded model and considering the period, gamma and roughness values of $W/B_4C$ and $B_4C/W$ interfaces are varying ( Figure~\ref{fig-clustergradedfit_sigmavarying_fitpar} ) in stacks. (a) N=170, (b) N=50.}
\label{fig-clustergradedfit_sigmavarying}
\end{figure*}

It is seen from the Figure \ref{fig-clustergradedfit_sigmacons}, that the measured reflectivity curve is well modeled by DarpanX except for a slight difference between the higher Bragg peaks.
These deviations may be because of the fact that the surface roughness values of $B_4C/W$ and $W/B_4C$- interfaces are not the same throughout the multilayer stack. 
In order to verify that, we keep the interlayer roughness values of each block of cluster-graded model as a free variable. 
Figure \ref{fig-clustergradedfit_sigmavarying} shows the fitted data. The variation of roughness, period and gamma with each block is shown in Figure \ref{fig-clustergradedfit_sigmacons_fitpar}.
The best-fit parameters that are common to all blocks are shown in the fifth (for N=170) and sixth (for N=50) rows of Table~\ref{panini_samples_table}.
\begin{figure*}
\centering
\begin{subfigure}{.5\textwidth}
  \centering
  \includegraphics[width=0.95\linewidth]{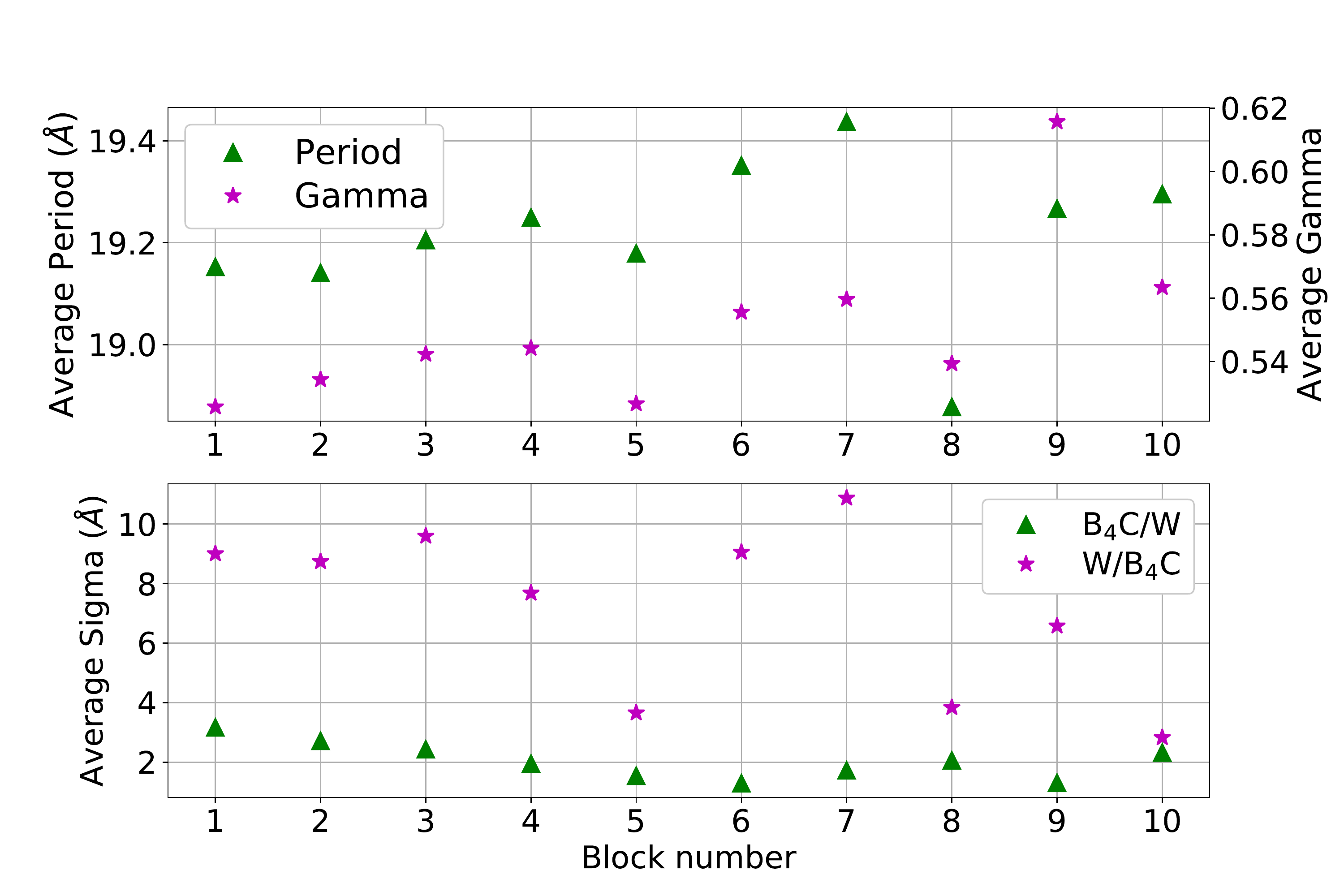} 
  \caption{}
  \label{fig:clustergradedfitd19_sigmavarying_fitpar}
\end{subfigure}%
\begin{subfigure}{.5\textwidth}
  \centering 
  \includegraphics[width=1.02\linewidth]{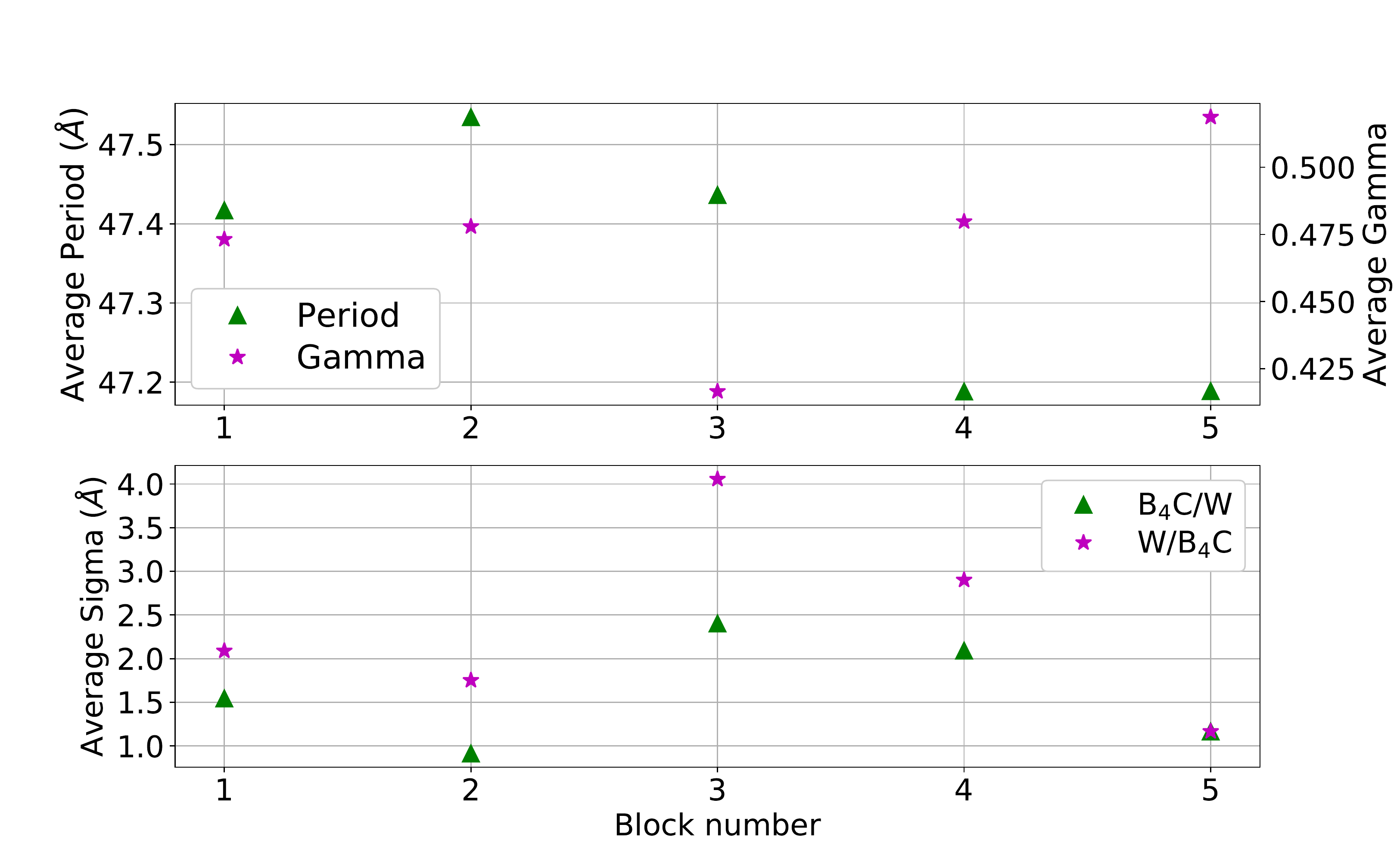}
  \caption{}
  \label{fig:clustergradedfitd44_sigmavarying_fitpar}
\end{subfigure}
 \caption{Variation of average $period$, $\Gamma$ of $W/B_4C$ bilayers and roughness values of $W/B_4C$ and $B_4C/W$ interfaces at each block of cluster-graded model for samples: (a) N=170 and (b) N=50.}
\label{fig-clustergradedfit_sigmavarying_fitpar}
\end{figure*}
 
The variation of period and gamma values within the different blocks of the cluster-graded model are less and it gives a better fit than the constant period bilayer model. 
Thus it can be seen that  XRR  data for the multilayer sample is well modeled with  DarpanX providing measurements of multiple parameters.

The capability of the DarpanX model in PyXspec to fit the X-ray reflectivity measurements of single-layer and multilayer samples to obtain the parameters of the system is established with these results. 
It also offers flexibility to carry out fitting with complex definitions of multilayer structure.

\section{Summary }\label{sec-summary}

We have initiated the development of the hard X-ray mirrors towards
the potential use in the future Indian X-ray astronomy mission and, in this context, a multilayer coating facility based on RF magnetron sputtering technique has been set up. 
In order to design multilayer mirrors and to characterize them using X-ray reflectivity (XRR) measurements, we have developed the DarpanX package that computes the reflectivity and other optical functions of multilayer systems. 
It can be used as a model for the X-ray fitting software 
XSPEC, which has robust fitting capabilities and thus allows
to estimations of various multilayer parameters. 
XSPEC has several physical models, which are generally used to fit the Astronomical data.
So far there are no models present within XSPEC to fit the XRR data. DarpanX adds this missing capability to XSPEC and thus makes XRR fitting/modeling much more widely available.
DarpanX has easy access to model the uncertain factor of the multilayer structure, such as cluster-graded or adding an extra top oxide layer/layers, etc.
The DarpanX algorithms are extensively tested for various types of multilayer structures.
It has been used to fit the experimentally
measured X-ray reflectivity data for both single and multilayer
samples and has been found to provide good fits.

DarpanX is developed in python3 and is packaged as a module that can be imported in interactive python shell or scripts.
The scripting capability makes it easy to optimize the multilayer design to suit for specific optical design or any other applicable constraints.  
Also, the parallel processing capability of DarpanX makes it faster in the case of large number of iteration required for the calculation.
DarpanX can be used as a stand-alone package to design any multilayer structure and estimate their physical properties.

DarpanX codes are publicly distributed and anybody can use/edit them or add different modules for other purposes. 
In the near future, it is planed to add the other module for the design of X-ray optics and estimate the different properties of it, such as the effective area of different types of optics, where the multilayer calculation will do by using the existing module and the other calculation (like geometrical calculation, etc.) will do by the new module. 
Also, an addition of a ray-tracing module to simulate the performance of the designed optics will be very helpful.
The multilayer design and characterization, using DarpanX, of the hard X-ray mirrors for the future astronomical telescope will be reported in subsequent publications.

\appendix 

\section{Theoretical Calculation of Reflectivity}\label{sec-our_code}
Laws of reflection/refraction at the boundary between two mediums for an electromagnetic wave are governed by the Fresnel's equations. These equations relate the amplitudes of reflected and refracted waves with those of the incident wave. Calculation of reflectivity in DarpanX is 
based on the Fresnel equations, modified for the finite surface roughness. 

\subsection{Fresnel equations for an ideal interface}
Consider an electromagnetic wave incident on the interface between two mediums whose refractive indices are $n_1$ and $n_2$ respectively. 
Let $\theta_1$ and $\theta_2$ be the incident and refracted angles (angles are measured from the interface). 
Then, Fresnel equations (eq:\ref{eq-f1}-\ref{eq-f4}), gives the amplitudes of reflected (r) and transmitted (t) waves\cite{born_wolf,jackson_classical_1999}. 

\begin{equation}\label{eq-f1}
r_\perp=\frac{n_1Sin(\theta_1) -n_2Sin(\theta_2)}{n_1Sin(\theta_1) +n_2Sin(\theta_2)}
\end{equation}

\begin{equation}\label{eq-f2}
r_\parallel=\frac{n_2Sin(\theta_1) -n_1Sin(\theta_2)}{n_2Sin(\theta_1) +n_1Sin(\theta_2)}
\end{equation}

\begin{equation}\label{eq-f3}
t_\perp=\frac{2n_1Sin(\theta_1)}{n_1Sin(\theta_1) +n_2Sin(\theta_2)}
\end{equation}

\begin{equation}\label{eq-f4}
t_\parallel=\frac{2n_1Sin(\theta_1)}{n_2Sin(\theta_1) +n_1Sin(\theta_2)}
\end{equation}
Here $\perp$ and $\parallel$ represent the perpendicular and parallel components of the electric field vectors with respect to the plane of interface of the incident wave. 
The square of amplitudes of reflected and transmitted waves given by Fresnel equations gives the optical functions, namely reflectivity and transmitivity.

\subsection{Optical functions for single/multi layer with ideal interface}\label{SLT}

Consider an electromagnetic wave of wavelength $\lambda$ is incident with an angle $\theta_0$ on a thin single layer film of thickness $d$ and refractive index $n_1$ . 
Then the incident rays undergo multiple partial reflections/refractions in between the two interface of the thin layer and a small fraction is returned to the initial medium. 
All of these components of rays produce an interference pattern.
Thus, considering the repeated reflection of rays in each interface and taking into account the phase difference ($\triangle \phi_{01}=\frac{4\pi}{\lambda}n_1 d.Sin(\theta_1)$) between the two successive reflected rays at the $1^{st}$ interface, and by summing all the components with applying energy conservation laws, gives the net amplitude of the rays reaching back to the initial medium as:
\begin{equation}\label{eq-sl2}
\Re=\frac{r_{01}+r_{12}e^{i\triangle\phi_{01}}}{1+r_{01}r_{12}e^{i\triangle\phi_{01}}}
\end{equation}
The physically measurable quantity is reflected intensity (reflectivity), $R={|\Re|}^2$. Similarly the transmission amplitude of the transmitted rays is given by\cite{born_wolf}, 
\begin{equation}\label{eq-sl3}
\tau=\frac{E_t}{E_0}=\frac{t_{01}t_{12}e^{\frac{i\triangle\phi_{01}}{2}}}{1+r_{01}r_{12}e^{i\triangle\phi_{01}}}
\end{equation}
Here $\theta_1$ is the refracted angle at $1^{st}$ interface of the layer and $n_0$ is the refractive index of ambient medium.
The transmitivity will be,  
\begin{equation}
T=|\tau|^2*\frac{n_1 Cos(\theta_1)}{n_0 Cos(\theta_0)}
\end{equation}

Multilayer mirrors consists of alternate high-Z (absorber) and low-Z (spacer) materials. Consider a plane wave incident on a multilayer, which is essentially a series of N layers located on a substrate. Then, the total number of interfaces
will be $N+1$. Let $\sigma_i$, $d_i$, and $n_i$ be the interfacial roughness, thickness and refractive index of the $i^{th}$ layer (where, i= 1,2,3....,N from the top layer) respectively and $n_0$, $n_s$ are the refractive indices of the ambient and substrate below the bottom most layers. Then from eq-\ref{eq-sl2}, the bottom most layer has amplitude of reflectivity,
\begin{equation}\label{eq-ml1}
\Re_N=\frac{r_{(N-1)N}+r_{N(N+1)}e^{i\triangle\phi_{(N-1)N}}}{1+r_{(N-1)N}r_{N(N+1)}e^{i\triangle\phi_{(N-1)N}}}
\end{equation}
\noindent
Here $(N+1)$ corresponds to the substrate and  $\triangle\phi_{(N-1)N}=\frac{4\pi}{\lambda}d_N Sin(\theta_{N})$ is the phase difference between the successive reflected rays from $N^{th}$ and $(N-1)^{th}$ interface.
Then the amplitude of reflection due to the combination of $N^{th}$ and $(N-1)^{th}$ layer will be,
\begin{equation}\label{eq-ml2}
\Re_{N-1}=\frac{r_{(N-2)(N-1)}+\Re_{N}e^{i\triangle\phi_{(N-2)(N-1)}}}{1+r_{(N-2)(N-1)}\Re_{N}e^{i\triangle\phi_{(N-2)(N-1)}}}
\end{equation}
\noindent
By progressively calculating the formula over all the $N$ layers of the multilayer structure, we get the final amplitude of reflection of the complete multilayer system as:
\begin{equation}\label{eq-ml3}
\Re_0=\frac{r_{01}+\Re_{1}e^{i\triangle\phi_{01}}}{1+r_{01}\Re_{1}e^{i\triangle\phi_{01}}}
\end{equation}
Then the reflectivity of the multilayer system for all the layers will be, $R=|\Re_0|^2$.
\noindent
Similarly, we can calculate the transmission amplitude and the transmittance of the multilayer system. Absorbance is given by 1-($R$ + $T$), for specular reflection .

\subsection{Interface Roughness Correction}
Above calculation is based on the consideration of ideal interface (i.e, surface roughness ($\sigma$) = 0). 
In reality, the boundary surfaces are rough.
In one dimension, the surface profile may be completely described by a function $z(x)$, which gives the height profile at every (x) points. For the best polished surface we can assume, 
$<z>=\frac{1}{L}\int_{0}^{L}  z(x) dx=0 \hspace{0.2cm} $ and  $\hspace{0.2cm} <z^2>=\frac{1}{L}\int_{0}^{L}  z^2(x) dx > 0 $.
Here $<z^2>=\sigma^2$ is called the variance and $\sigma$ is called the rms surface roughness and $L$ is the length of the surface along $x$-direction. 
For a smooth surface $z(x)=z_0=constant$, which does not hold true for real surfaces. 
Thus the roughness of a surface can characterize by its rms value $\sigma$.

Suppose $w(z)$ is the height distribution of $z(x)$, then the Fresnel coefficients can be modified by multiplying the
Debye-Waller factor\cite{spiga_thesis}, as:
\begin{equation}\label{eq-deby}
r'[i]=r[i] *\bar{w}(q_z[i])
\end{equation}
\noindent
Here $r[i]$ is the Fresnels coefficient for smooth surface and $\bar{w}(q_z[i])=\int w(z)exp(izq_z[i]) dz $ where, $q_z[i]=\frac{2\pi}{\lambda}sin\theta_i$ and $\theta_i$ is the propagation angle in the $i^{th}$ layer.

For the low spatial frequencies of the roughness spectrum, the Fresnel reflection coefficient
is usually multiplied by the Debye-Waller factor (equation \ref{eq-deby}), while for the high spatial frequencies, the correction coefficient is given by the Nevot-Croce correction factor \cite{surfaceroughness1,ref-novot_corr,imd_manual}. 
This model is widely used for the calculation of the X-ray reflectivity of multilayers. 
Considering Nevot-Croce correction factor one can modify the Fresnel's coefficients by using the following equation:

\begin{equation}\label{eq-41}
\hspace{0.5cm} r'[i]=r[i]*\bar{w}(2\sqrt{q_z[i].q_z[i+1]})
\end{equation}
\noindent

Stearns et.al\cite{ref8} provided the functional form of `$w$' and corresponding `$\bar{w}$' for different type (error function, exponential, linear, sinusoidal) of one dimensional interface profiles. 
These functional forms can be use for different type of interface profiles. 


\bibliographystyle{model3-num-names}

\bibliography{myref}

\end{document}